\documentclass[review,12pt,sort&compress]{elsarticle}
\usepackage[utf8]{inputenc}
\usepackage{siunitx}
\usepackage{geometry}
\usepackage{comment}
\usepackage{epsfig}
\makeatletter \def\ps@pprintTitle{  \let\@oddhead\@empty  \let\@evenhead\@empty  \def\@oddfoot{\small{\today\hfill\thepage}}  
\def\@evenfoot{\thepage\hfill}} 
\makeatother
\usepackage{amsfonts,amssymb,amsmath,amsthm,amsbsy}
\usepackage{graphicx}
\usepackage[T1]{fontenc}
\usepackage[utf8]{inputenc}
\usepackage{listings}
\usepackage{appendix}
\usepackage{minted}
\usepackage{lineno}
\usepackage{color}
\definecolor{mygreen}{RGB}{28,172,0}
\definecolor{mylilas}{RGB}{170,55,241}
\usepackage{mathtools}
\numberwithin{equation}{section}
\usepackage{caption}
\usepackage{subcaption}
\usepackage{subfiles}
\usepackage{float}
\usepackage[colorlinks=true,citecolor=blue,urlcolor=blue]{hyperref}
\usepackage{silence}
\usepackage{amsmath}
\numberwithin{equation}{section}
\usepackage{pgfplots}
\usepackage{tikz}
\pgfplotsset{compat=1.18}

\begin{document}

% ==============================================================
% Frontmatter
% ============================================================== 
\begin{frontmatter}

\title{Flow of a colloidal solution in an orthogonal rheometer}

\author{Krishna Kaushik Yanamundra, Chandler C. Benjamin, Kumbakonam Ramamani Rajagopal}
\address{Department of Mechanical Engineering, Texas A\&M University, College Station, TX 77843, USA}

\begin{abstract}

The flow of a colloidal solution between two parallel disks rotating with the same angular velocity about two non-coincident axes was studied. The problem has been approached from two perspectives, the first wherein the stress is expressed in terms of a power-law of kinematical quantities, and the second wherein we consider a non-standard model where the symmetric part of the velocity gradient is given by a power-law of the stress. For a range of power-law exponents, the class of models are non-invertible. By varying the material and geometric parameters, changes in the flow behaviour at different Reynolds numbers were analysed. We find that pronounced boundary layers develop even at low Reynolds numbers based on the power-law exponents. The new class of stress power-law fluids and fluids that exhibit limiting stress also show the ability to develop boundary layers.

\end{abstract}

\begin{keyword}

\texttt{Orthogonal rheometer \sep stress power-law fluids \sep implicit constitutive relation \sep non-Newtonian fluids \sep boundary layers \sep colloidal solutions \sep constitutive relations for colloids.}
\end{keyword}
\end{frontmatter}

\section{\label{sec:Introduction}Introduction}

The study of fluid flow due to rotating disks has been the subject of numerous classical studies and continues to receive considerable scholarly attention. This includes the flow due to the rotation of a single disk, starting with the seminal study of von Karman \cite{karman1921laminare} followed by numerous others (see the review article by Zandbergen and Dijkstra \cite{zandbergen1987karman}), as well as the flow between two parallel disks, either rotating about a common axis \cite{batchelor1951note, stewartson1953flow, berker1979new} or distinct axes with the same angular velocity \cite{abbott_walters_1970, rajagopal1982flow}. A comprehensive discussion of flows between rotating parallel plates can be found in the review article by Rajagopal \cite{rajagopal1992flow}. \\
The problem of flow between eccentrically rotating disks has attracted considerable attention owing to its relevance to the flow occurring in the ``orthogonal rheometer" developed by Maxwell and Chartoff \cite{maxwell1965studies}. Bird and Harris \cite{bird1968analysis}, Blyler and Kurtz \cite{blyler1967analysis}, Kearsley \cite{kearsley1970flow}, and Gordon and Schowalter \cite{gordon1970relation} carried out early studies in such a flow domain. An orthogonal rheometer is essentially two parallel plates that rotate about two distinct non-coincident axes with the same angular velocity; the fluid to be tested fills the space between them. The instrument is capable of measuring forces in three orthogonal directions. Hence, it is particularly useful in characterizing the material properties of non-Newtonian fluids that develop normal stress differences in shear flows. Assuming that the plates are infinite and that the edge effects can be neglected, and further assuming that the flow taking place in such an instrument has a special form, the flows of several fluids have been studied in the past \cite{goldstein1975further, huilgol1969properties,ahrens1977viscous,rajagopal1981flow,rajagopal1986flow,bower1987flow,rajagopal1983flow,ersoy1999mhd,al2003flow,ravindran2004note,tigoiu2000flow}. Most of these studies are restricted to the class of simple fluids (see Truesdell and Noll \cite{truesdell2004non}) wherein the stress is a functional of the history of the relative deformation gradient.\footnote{Since the stress (traction) is the cause and motion (kinematics) the effect, prescribing the stress as a function of kinematical quantities is contrary to causality.} Moreover, there are several materials, especially colloids, whose behavior cannot be described by the constitutive relation of a simple fluid or for the rate type fluids that are commonly used (see Boltenhagen et al. \cite{boltenhagen1997observation}). In view of this, Rajagopal \cite{rajagopal2006implicit} generalized the class of constitutive relations to a functional relationship between the history of the stress and the history of the relative deformation gradient. A special sub-class of these, fluids described by an implicit algebraic constitutive relation (see Perlocova and Prusa \cite{perlacova2015tensorial}, Rajagopal \cite{rajagopal2016flows}) seems to be adequate to describe the behavior of colloids and suspensions that show non-monotonic stress-strain rate relations.
\par Malek et al. \cite{malek2010generalizations} introduced one such constitutive relation wherein the symmetric part of the velocity gradient was given by a “power-law” of the deviator of the stress. They obtained exact analytical solutions for this model undergoing plane Couette, plane Poiseuille, Hagen-Poiseuille, and cylindrical Couette flows. In the case of cylindrical Couette flow, the boundary conditions were found to have a significant impact on the existence and uniqueness of the solution. Srinivasan and Karra \cite{srinivasan2015flow} studied this model in the context of flow due to eccentrically rotating disks. They observed a similar effect of the boundary conditions on the problem's well-posedness and found that the system of governing equations admits either one, two, or no solutions depending on the parameter values, when inertial effects are neglected.
\par A modification to this stress power-law model was proposed by Le Roux and Rajagopal \cite{le2013shear} to better model fluids that show an S-type non-monotonic relation between strain-rate and stress, as observed in the results presented by Boltenhagen et al. \cite{boltenhagen1997observation}. Fusi et al. \cite{fusi2023flow} studied the flow of this modified stress power-law model in an orthogonal rheometer where they observed the development of boundary layers even at lower Reynolds numbers in contrast to the inertial boundary layers one would observe at high Reynolds numbers in the case of a linearly viscous fluid. 
\par This paper proposes a further generalization of the stress-power law model that allows us to model fluids which exhibit ``stress-limiting" behaviour and also those fluids whose fluidity tends to infinity when subjected to small stresses but behaves like a linearly viscous fluid as stress increases. We study the flow of this ``generalized stress power-law fluid" and also of its counterpart the generalized classical power-law model in an orthogonal rheometer. We perform a parametric analysis on both these models to study how the flow is altered as the material and flow parameters vary.
\section{Constitutive relations}
\begin{figure}[H]
    \centering    
    \captionsetup{width=\linewidth}
    \scalebox{0.75}{
    \includegraphics[]{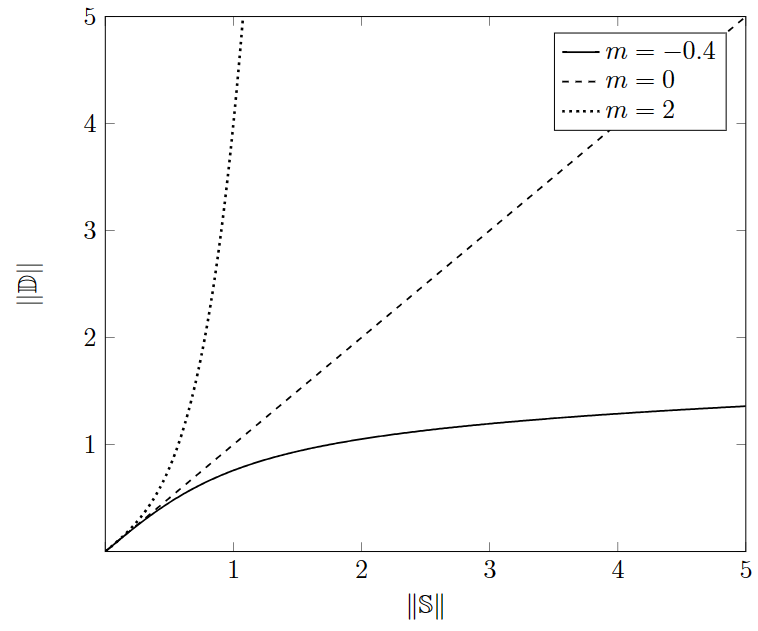}}
\caption{Behaviour of (\ref{Eqn1}) for some values of  $m$ ($\alpha=1$, $a_1=0$, $a_2=1$, $\beta=1$)}
\label{Fig1}
\end{figure}

 We study the flow of incompressilbe fluids whose Cauchy stress tensor is given by -
 \begin{align}
 \mathbb{T}:= -p\mathbb{I} + \mathbb{S}    
 \end{align}
 where $-p\mathbb{I}$ is the indeterminate part of the stress due to the constraint of incompressibility, and  $\mathbb{S}$ is the deviatoric part of the stress tensor. While in a simple fluid, the deviatoric part is expressed as function of the symmetric part of the velocity gradient, $\mathbb{D}$, in the generalized stress-power law model (\ref{Eqn1}), we express $\mathbb{D}$ as a function of $\mathbb{S}$:
\begin{align}
    \mathbb{D} &= \alpha \left(a_1 + \left(a_2 + \beta \|\mathbb{S}\|^2\right)^m\right)\mathbb{S} \label{Eqn1}\\
    \mathbb{S} &= \mu \left(b_1 + \left(b_2 + \gamma \|\mathbb{D}\|^2\right)^n\right)\mathbb{D}
    \label{Eqn2}
\end{align}
$\|\mathbb{A}\|$ denotes the usual trace norm of a tensor $\mathbb{A}$ and is given by $\|\mathbb{A}\|^2 = \text{Tr}(\mathbb{A}\mathbb{A}^T)$. For the models to be thermodynamically admissible, we require that the generalized fluidity in (\ref{Eqn1}) and the generalized viscosity in (\ref{Eqn2}) be non-negative. Therefore in (\ref{Eqn1}),
\begin{align}
\alpha \geq 0,\mbox{ } a_1 \geq 0, \mbox{ and } \beta \geq 0    
\end{align}
and similarly in (\ref{Eqn2}),
\begin{align}
\mu \geq 0,\mbox{ } b_1 \geq 0, \mbox{ and } \gamma \geq 0    
\end{align}
Without loss of generality, we can restrict $a_2$ and $b_2$ to be 0 or 1. For model (\ref{Eqn1}) to be meaningful as $\|\mathbb{S}\| \rightarrow 0$, 
\begin{align}
      m \in \begin{cases}
        \left[-\frac{1}{2},\infty\right), &\text{$a_2=0$}
        \\
        \left(-\infty,\infty\right), &\text{$a_2=1$}
        \end{cases}
\end{align}
and similarly for model (\ref{Eqn2}) to be meaningful as $\|\mathbb{D}\| \rightarrow 0$, 
\begin{align}
      n \in \begin{cases}
        \left[-\frac{1}{2},\infty\right), &\text{$b_2=0$}
        \\
        \left(-\infty,\infty\right), &\text{$b_2=1$}
        \end{cases}
\end{align}
Let us qualitatively analyze each of these models for different ranges of parametric values. For $a_2 = 1$ as $m$ varies the generalized stress power-law model behaves as follows -
\begin{align}
      m \in \begin{cases}
        \left[-\frac{1}{2},0\right): &\text{(\ref{Eqn1}) exhibits stress-thickening}\\
        \{0\}: &\text{(\ref{Eqn1}) reduces to a linearly viscous model}\\
        \left(0,\infty\right): &\text{(\ref{Eqn1}) exhibits stress-thinning}
        \end{cases}
\end{align}
This dependence of (\ref{Eqn1}) on $m$ has been illustrated in Figure \ref{Fig1}.

\begin{figure}[H]
    \centering    
    \captionsetup{width=\linewidth}
    \scalebox{0.75}{
    \includegraphics[]{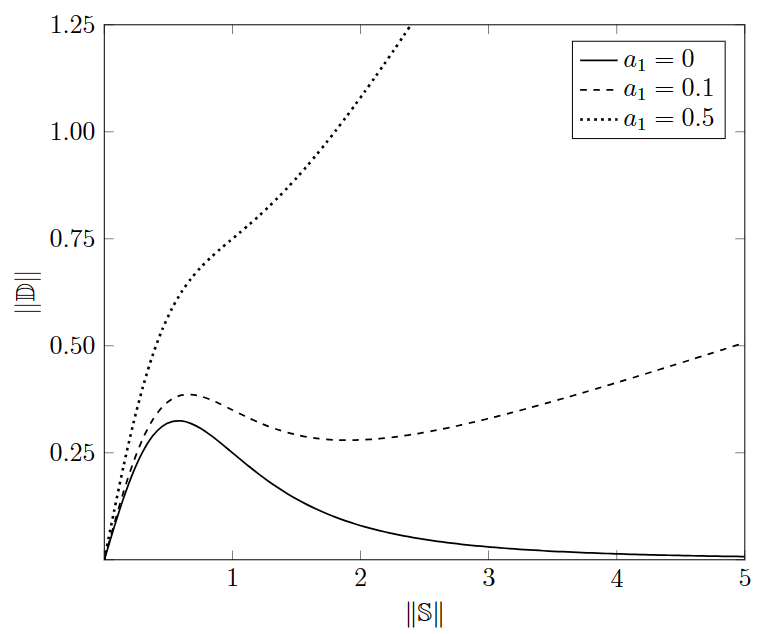}}
\caption{Behaviour of (\ref{Eqn1}) for some values of  $a_1$ ($\alpha=1$, $n=-2$, $a_2=1$, $\beta=1$)}
\label{Fig2}
\end{figure}

When $a_2 = 1$ and $m<-\frac{1}{2}$, nature of (\ref{Eqn1}) depends on $a_1$ as follows (Refer to Le Roux and Rajagopal \cite{le2013shear} for proof) -
\begin{align}
      a_1 \in \begin{cases}
        \{0\}: &\text{non-monotonic with one inflection point}\\
        \left(0,2\left(\frac{\left|2m+1\right|}{2\left(1-m\right)}\right)^{\left(1-m\right)}\right): &\text{S-type non-monotonicity with two inflection points}\\
        \left[2\left(\frac{\left|2m+1\right|}{2\left(1-m\right)}\right)^{\left(1-m\right)},\infty\right): &\text{monotonically increasing}
        \end{cases}
\end{align}
This dependence of (\ref{Eqn1}) on $a_1$, when $m<-\frac{1}{2}$, has been illustrated in Figure \ref{Fig2}.

For $a_2 = 0$ and $m<0$, the generalized fluidity, $\alpha_g(\|\mathbb{S}\|)$ given by $\alpha \left(a_1 + \left(a_2 + \beta \|\mathbb{S}\|^2\right)^m\right)$ in (\ref{Eqn1}) tends to $\infty$ as $\|\mathbb{S}\| \rightarrow 0$ behaving like an Euler fluid and asymptotically reaches $a_1\alpha$ as $\|\mathbb{S}\| \rightarrow \infty$, akin to a linear model. As $m \rightarrow \infty$, the model shows ``stress-limiting" behaviour. Dependence of $\alpha_g(\|\mathbb{S}\|)$ on $m$ when $a_2 = 0$ has been illustrated using Figure \ref{Fig3} below.

\begin{figure}[H]
    \centering    
    \captionsetup{width=\linewidth}
    \scalebox{0.75}{
    \includegraphics[]{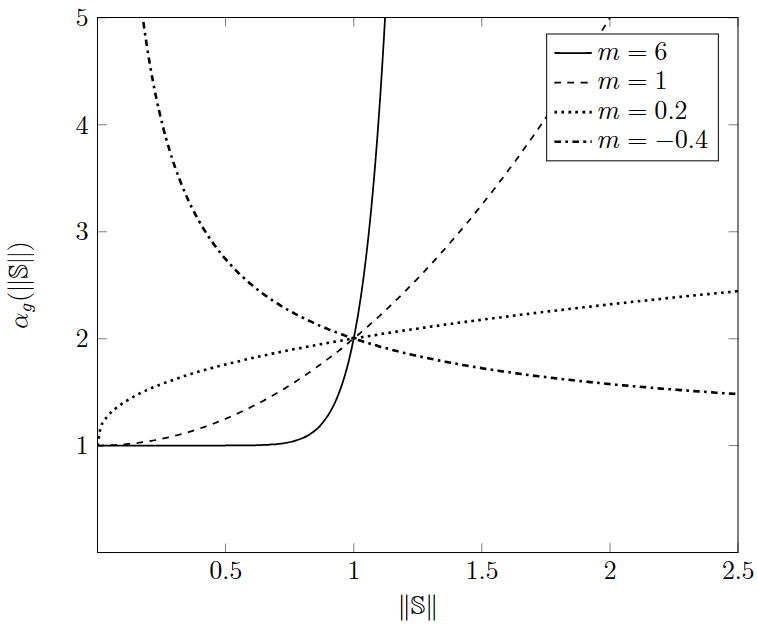}}
\caption{Behaviour of $\alpha_g(\|\mathbb{S}\|)$ for (\ref{Eqn1}) for some values of  $m$ ($\alpha=1$, $a_1=1$, $a_2=0$, $\beta=1$)}
\label{Fig3}
\end{figure}

Only when $a_1 = 0$ and $a_2 =0$, we can obtain a closed form inverse in the form of model (\ref{Eqn2}), with -
\begin{align}
&b_1 = 0,& &b_2 = 0,& &\mu = \alpha^{\frac{-1}{2m +1}},& &\gamma = \beta,& &\text{and}& &n = \frac{-m}{2m+1}&
\end{align}
The generalized classical power-law model (\ref{Eqn2}), when $b_2 = 1$, and as $n$ varies, behaves as follows - 
\begin{align}
      n \in \begin{cases}
        \left[-\frac{1}{2},0\right): &\text{(\ref{Eqn2}) exhibits shear-thinning}\\
        \{0\}: &\text{(\ref{Eqn2}) reduces to a linearly viscous model}\\
        \left(0,\infty\right): &\text{(\ref{Eqn2}) exhibits shear-thickening}
        \end{cases}
\end{align}

When $b_2 = 0$, and $n<0$, the generalized viscosity of (\ref{Eqn2}) tends to $\infty$ as $\|\mathbb{D}\| \rightarrow 0$ which means the resistance to flow increases, which is akin to a ``yield-stress" coming into play. On the other hand, when $n \rightarrow \infty$, the model exhibits shear-limiting behaviour.

\section{Boundary value problem}
\begin{figure}[H]
    \centering
    \includegraphics[scale=0.6]{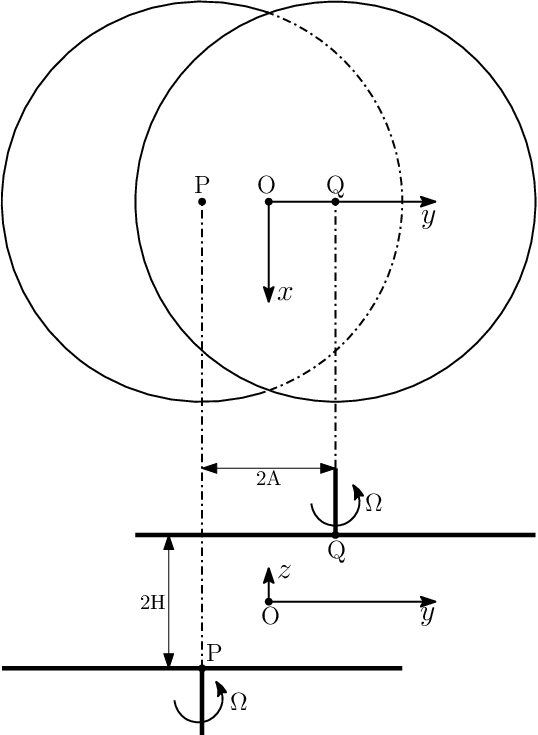}
    \caption{Schematic diagram of an orthogonal rheometer}
    \label{Fig:OR}
\end{figure}

Let's explore the fluid flow governed by the constitutive relations described in equations (\ref{Eqn1}) and (\ref{Eqn2}) between two parallel disks with infinite radii, positioned with their centers 2A units apart and a gap of 2H between them. The origin is symmetrically situated, aligning the centers of rotation of the top and bottom disks at coordinates (0, A, H) and (0, -A, -H) respectively. Both disks are assumed to rotate at a constant angular velocity $\Omega$. This setup is illustrated in Figure \ref{Fig:OR}.

Consider the flow given through (\ref{Eqn3.1}). where $u$, $v$, and $w$ represent the components of the velocity field $\textbf{v}$ in the $x$, $y$, and $z$ directions, respectively.

\begin{align} \label{Eqn3.1}
u &= -\Omega(y - g(z))\nonumber\\ 
v &= \Omega(x - f(z))\\ 
w &= 0\nonumber
\end{align}

The above flow belongs to the class of pseudo-planar motions of 1st kind as delineated by Berker \cite{berker1936quelques}. The fluid in each plane rotates about a centre with some constant angular velocity. $f(z)$ and $g(z)$ signify the projections of the locus of these centres of rotation (also referred to as the stagnation points) on the $xz$ and $yz$ planes respectively. It was shown by Rajagopal \cite{rajagopal1982flow} that this motion is one with constant principal relative stretch history of 3rd kind (also referred to as nearly-viscometric flow), a classification introduced by Noll \cite{noll1962motions}. Later, Goddard \cite{goddard1983dynamics} independently obtained the same result. From (\ref{Eqn3.1}), it follows that the velocity gradient $\mathbb{L}$ given by  (\ref{Eqn3.2})%\footnote{The notation $[\mathbb{L}]$ is used to denote the representation of the tensor $\mathbb{L}$ in the matrix form.}.
\begin{equation} \label{Eqn3.2}
\mathbb{L} = \frac{\partial \textbf{v}}{\partial \textbf{x}} = \Omega\begin{bmatrix}
0 & -1& \frac{dg}{dz}\\
1 & 0 & -\frac{df}{dz}\\
0 & 0 & 0
\end{bmatrix}
\end{equation}
From $\mathbb{L}$, we can find the symmetric part of the velocity gradient $\mathbb{D}$ and its norm-
\begin{equation}\label{Eqn3.3}
\mathbb{D} = \frac{1}{2}(\mathbb{L} + \mathbb{L}^T) = \frac{\Omega}{2}\begin{bmatrix}
0 & 0& \frac{dg}{dz}\\
0 & 0 & -\frac{df}{dz}\\
\frac{dg}{dz} & -\frac{df}{dz} & 0
\end{bmatrix}
\end{equation}
\begin{equation}\label{Eqn3.4}
\|\mathbb{D}\|^2 = \text{Tr}(\mathbb{D}\mathbb{D}^T) = \frac{\Omega^2}{2}\left(\frac{dg}{dz}^2 + \frac{df}{dz}^2\right)
\end{equation}
And one could also compute the skew-symmetric part of the velocity gradient $\mathbb{W}$ and its norm-
\begin{equation}\label{Eqn3.5}
\mathbb{W} = \frac{1}{2}(\mathbb{L} - \mathbb{L}^T) = \frac{\Omega}{2}\begin{bmatrix}
0 & -2& \frac{dg}{dz}\\
2 & 0 & -\frac{df}{dz}\\
-\frac{dg}{dz} & \frac{df}{dz} & 0
\end{bmatrix}
\end{equation}

\begin{equation}\label{Eqn3.6}
\|\mathbb{W}\|^2 = \text{Tr}(\mathbb{W}\mathbb{W}^T) = \frac{\Omega^2}{2}\left(\frac{dg}{dz}^2 + \frac{df}{dz}^2 + 4\right)
\end{equation}
By virtue of our velocity ansatz (\ref{Eqn3.1}), we have the following form for the deviator of the Cauchy stress $\mathbb{T}$.
\begin{subequations}  \label{Eqn3.7}
\begin{align}
\mathbb{T} &= -p\mathbb{I} + \mathbb{S} \label{Eqn3.7a}\\
\mathbb{S} &= \begin{bmatrix}
0 & 0 & S_{13}\\
0 & 0 & S_{23}\\
S_{13} & S_{23} & 0
\end{bmatrix} \label{Eqn3.7b}\\
\|\mathbb{S}\|^2 &= \text{Tr}(\mathbb{S}\mathbb{S}^T) = 2\left(S_{13}^2 + S_{23}^2\right) \label{Eqn3.7c}
\end{align}
\end{subequations}
Now, from the balance of linear momentum and (\ref{Eqn3.7a}), we have - 
\begin{subequations} \label{Eqn3.8}
\begin{align}
\rho \frac{d\textbf{v}}{dt} &= div(\mathbb{T}) + \rho\textbf{b}\\
\implies \rho \frac{d\textbf{v}}{dt} &= -\frac{\partial p}{\partial \textbf{x}} + div(\mathbb{S}) + \rho\textbf{b} \label{Eqn3.8b}
\end{align}
\end{subequations}
where $p$ is the Lagrange multiplier due to the constraint of incompressibility, and $\textbf{b}$ is the body force. From (\ref{Eqn3.1}) we get - 

\begin{equation} \label{Eqn3.9}
\frac{d\textbf{v}}{dt} = -\Omega^2 \begin{bmatrix}
(x - f(z))\\
(y - g(z))\\
0
\end{bmatrix}
\end{equation}
and from (\ref{Eqn3.7b}) we obtain -
\begin{equation} \label{Eqn3.10}
div(\mathbb{S}) = \begin{bmatrix}
\frac{dS_{13}}{dz}\\
\frac{dS_{23}}{dz}\\
0
\end{bmatrix}
\end{equation}
Assuming that the body forces are neglected, after substituting (\ref{Eqn3.9}) and (\ref{Eqn3.10}) in (\ref{Eqn3.8b}) we get the following -
\begin{subequations}
\begin{align}
\rho\Omega^2f(z) &= -\frac{\partial p}{\partial x} + \rho\Omega^2x + \frac{dS_{13}}{dz}\label{Eqn3.12a}\\
\rho\Omega^2g(z) &= -\frac{\partial p}{\partial y} + \rho\Omega^2y + \frac{dS_{23}}{dz}\label{Eqn3.12b}\\
\frac{\partial p}{\partial z} &= 0 \label{Eqn3.12c}
\end{align}
\end{subequations}
From (\ref{Eqn3.12c}) we realize that $p$ is only a function of $x$ and $y$. Then from (\ref{Eqn3.12a}) and (\ref{Eqn3.12b}), we obtain the following functional form for $p$.
\begin{equation}\label{Eqn3.14}
p(x,y) = -  \frac{\rho\Omega^2}{2}\left(x^2 + y^2\right) + c_1x + c_2y + c_3
\end{equation}
It can be noticed that the functional form for $p(x,y)$ in (\ref{Eqn3.14}) that the terms $c_1x$ and $c_2y$ would lead to Poiseuille flows in the x and y directions. In order to eliminate such flows, we shall seek solutions with $c_1 = 0$ and $c_2 = 0$.
We get 2 scalar equations after substituting (\ref{Eqn3.14}) in (\ref{Eqn3.12a}) and (\ref{Eqn3.12b}). These conditions are to be met for both constitutive models (\ref{Eqn1}) and (\ref{Eqn2}).
\begin{subequations}
\begin{align}
&\frac{dS_{13}}{dz} = \rho \Omega^2 f(z)\label{Eqn3.15a}\\
&\frac{dS_{23}}{dz} = \rho \Omega^2 g(z)\label{Eqn3.15b}
\end{align}
\end{subequations}
Now for the case of the generalized stress power-law model by substituting (\ref{Eqn3.3}), (\ref{Eqn3.7b}) and (\ref{Eqn3.7c}) in (\ref{Eqn1}), we get - 
\begin{subequations}
\begin{align}
&\frac{dg}{dz} = \frac{2\alpha}{\Omega}\left(a_1 + \left(a_2 + 2\beta\left(S_{13}^2 + S_{23}^2\right)\right)^m\right)S_{13} \label{Eqn3.16a}\\
&\frac{df}{dz} = -\frac{2\alpha}{\Omega}\left(a_1 + \left(a_2 + 2\beta\left(S_{13}^2 + S_{23}^2\right)\right)^m\right)S_{23} \label{Eqn3.16b}
\end{align}
\end{subequations}
Similarly, for its classical counterpart, by substituting (\ref{Eqn3.3}), (\ref{Eqn3.4}) and (\ref{Eqn3.7b}) in (\ref{Eqn2}), we get - 
\begin{subequations}
\begin{align}
&S_{13} = \frac{\Omega \mu }{2}\left(b_1 + \left(b_2 + \frac{\gamma\Omega^2}{2}\left(\frac{dg}{dz}^2 + \frac{df}{dz}^2\right)\right)^n\right)\frac{dg}{dz} \label{Eqn3.17a}\\
&S_{23} = -\frac{\Omega \mu }{2}\left(b_1 + \left(b_2 + \frac{\gamma\Omega^2}{2}\left(\frac{dg}{dz}^2 + \frac{df}{dz}^2\right)\right)^n\right)\frac{df}{dz} \label{Eqn3.17b}
\end{align}
\end{subequations}
For the problem to be well-posed, we need to supplement both systems of equations with appropriate boundary conditions that describe the flow at the surface of the rotating plates. If we assume no-slip condition at both the plates, we can obtain the velocity of the fluid at the top and bottom plates as follows -   
\begin{align} \label{Eqn}
u(x,y,\pm H) &= -\Omega(y \mp A)\nonumber\\ 
v(x,y,\pm H) &= \Omega x\\ 
w(x,y,\pm H) &= 0\nonumber
\end{align}
On comparing (\ref{Eqn}) and (\ref{Eqn3.1}) we then obtain the boundary conditions on $f$ and $g$ as - $f(\pm H) = 0$ and $g(\pm H) = \pm A$. Once we arrive at a solution, we can go further and compute the tractions on the plates.
\begin{equation}
    \textbf{t} = \mathbb{T}^T\textbf{n}
\end{equation}
where, $\textbf{t}$ is the traction vector and $\textbf{n}$ is the unit outward normal vector to the surface of the plate. Using this we can compute the moment acting on the plates.
\begin{equation}
    \textbf{m} = \int(\textbf{x}\wedge\mathbb{T}^T\textbf{n})~da
\end{equation}
where $\textbf{m}$ is the moment and $\textbf{x}$ is the position vector. It has been shown by Srinivasan and Karra \cite{srinivasan2015flow} that the net moment will be zero once the steady state is achieved due to the assumed ansatz, irrespective of the constitutive relation. However, we can always compute the local contribution at the centre of the plate to the resultant moment as follows - 
\begin{align}
m_x = yt_z - zt_y\\
m_y = zt_x - xt_z\\
m_z = xt_y - yt_x \label{Eqn3.22}
\end{align}
Since we are only interested in the moment about the $z$-axis, upon substituting for the coordinates of the centre of the top plate in (\ref{Eqn3.22}) we get -
\begin{equation}
    m_z = -AS_{13}
\end{equation}
\section{Non-dimensionalization}
To non-dimensionalize the set of ODEs we have obtained, we use the following quantities. For model (\ref{Eqn1}), 
 
\begin{align*}
&\overline{z} = \frac{z}{H}&&\overline{f(z)} = \frac{f(z)}{A}& &\overline{g(z)} = \frac{g(z)}{A}\\
&\overline{S_{13}} = \frac{S_{13}\alpha}{\Omega}& &\overline{S_{23}} = \frac{S_{23}\alpha}{\Omega}& &\overline{\beta} = \frac{\beta\Omega^2}{\alpha^2}\\
&\epsilon = \frac{A}{H}&&R = \rho \Omega H^2\alpha&
\end{align*}
The equations that are to be solved for model (\ref{Eqn1}) after non-dimensionalization are - 
\begin{subequations}
\begin{align}
&\frac{d\overline{f}}{d\overline{z}} = -\frac{2}{\epsilon}\left(a_1 + \left(a_2 + 2\overline{\beta}\left(\overline{S_{13}}^2 + \overline{S_{23}}^2\right)\right)^m\right)\overline{S_{23}}\label{Eqn4.1a}\\
&\frac{d\overline{g}}{d\overline{z}} = \frac{2}{\epsilon}\left(a_1 + \left(a_2 + 2\overline{\beta}\left(\overline{S_{13}}^2 + \overline{S_{23}}^2\right)\right)^m\right)\overline{S_{13}}\label{Eqn4.1b}\\
&\overline{f} = \frac{1}{\epsilon R}\frac{d\overline{S_{13}}}{d\overline{z}}\label{Eqn4.1c}\\
&\overline{g} = \frac{1}{\epsilon R}\frac{d\overline{S_{23}}}{d\overline{z}}\label{Eqn4.1d}\\
&\overline{f}(\pm1) = 0\\
&\overline{g}(\pm1) = \pm1
\end{align}
\end{subequations}
By using (\ref{Eqn4.1c}) and (\ref{Eqn4.1d}) to eliminate $\overline{f}$ and $\overline{g}$ from the above system of equations, we can obtain a system of two second-order ODEs on $\overline{S_{13}}$ and $\overline{S_{23}}$ with four Neumann boundary conditions for the components of stresses.

\begin{subequations}\label{Eqn4.2}
\begin{align} 
&\frac{d^2\overline{S_{13}}}{d\overline{z}^2} = -2R\left(a_1 + \left(a_2 + 2\overline{\beta}\left(\overline{S_{13}}^2 + \overline{S_{23}}^2\right)\right)^m\right)\overline{S_{23}}\\
&\frac{d^2\overline{S_{23}}}{d\overline{z}^2} = 2R\left(a_1 + \left(a_2 + 2\overline{\beta}\left(\overline{S_{13}}^2 + \overline{S_{23}}^2\right)\right)^m\right)\overline{S_{13}}\\
&\frac{d\overline{S_{13}}}{d\overline{z}}(\pm1) = 0\\
&\frac{d\overline{S_{23}}}{d\overline{z}}(\pm1) = \pm\epsilon R
\end{align}
\end{subequations}

Similarly for the classical model (\ref{Eqn2}), we can use the following non-dimensionalization.
\begin{align*}
&\overline{z} = \frac{z}{H}&&\overline{f(z)} = \frac{f(z)}{A}& &\overline{g(z)} = \frac{g(z)}{A}\\
&\overline{S_{13}} = \frac{S_{13}}{\mu\Omega}& &\overline{S_{23}} = \frac{S_{23}}{\mu\Omega}& &\overline{\gamma} = \gamma\Omega^2\\
&\epsilon = \frac{A}{H}&&R = \frac{\rho \Omega H^2}{\mu}&
\end{align*}

The system of ODEs to be solved for model (\ref{Eqn2}) after non-dimensionalization is as follows - 
\begin{subequations}
\begin{align}\label{Eqn4.3}
&\overline{S_{13}} = \frac{\epsilon}{2}\left(b_1 + \left(b_2+ \frac{\epsilon^2 \overline{\gamma}}{2}\left(\frac{d\overline{g}}{d\overline{z}}^2 + \frac{d\overline{f}}{d\overline{z}}^2\right)\right)^n\right)\frac{d\overline{g}}{d\overline{z}}\\
&\overline{S_{23}} = -\frac{\epsilon}{2}\left(b_1 + \left(b_2 + \frac{\epsilon^2 \overline{\gamma}}{2}\left(\frac{d\overline{g}}{d\overline{z}}^2 + \frac{d\overline{f}}{d\overline{z}}^2\right)\right)^n\right)\frac{d\overline{f}}{d\overline{z}}\\
&\frac{d\overline{S_{13}}}{d\overline{z}} = \epsilon R \overline{f}\\
&\frac{d\overline{S_{23}}}{d\overline{z}} = \epsilon R \overline{g}\\
&\overline{f}(\pm1) = 0\\
&\overline{g}(\pm1) = \pm1
\end{align}
\end{subequations}

We can eliminate $\overline{S_{13}}$ and $\overline{S_{23}}$ from the above system (\ref{Eqn4.3}) of equations to obtain a system of two second-order ODEs in $\overline{f}$ and $\overline{g}$.

\begin{subequations}\label{Eqn4.4}
\begin{align}
&\frac{d}{d\overline{z}}\left(\left(b_1 + \left(b_2+ \frac{\epsilon^2 \overline{\gamma}}{2}\left(\frac{d\overline{g}}{d\overline{z}}^2 + \frac{d\overline{f}}{d\overline{z}}^2\right)\right)^n\right)\frac{d\overline{g}}{d\overline{z}}\right) = 2R\overline{f}\\
&\frac{d}{d\overline{z}}\left(\left(b_1 + \left(b_2 + \frac{\epsilon^2 \overline{\gamma}}{2}\left(\frac{d\overline{g}}{d\overline{z}}^2 + \frac{d\overline{f}}{d\overline{z}}^2\right)\right)^n\right)\frac{d\overline{f}}{d\overline{z}}\right) = -2R\overline{g}\\
&\overline{f}(\pm1) = 0\\
&\overline{g}(\pm1) = \pm1
\end{align}
\end{subequations}

\section{Results}
The coupled systems of nonlinear ordinary differential equations (\ref{Eqn4.2}) and (\ref{Eqn4.4}) were numerically solved using the general form PDE interface within the mathematical module of COMSOL Multiphysics version 5.5. The nonlinear system of ODEs is discretized using Hermite cubic elements and Newton's method is used for solving the resulting discrete system, along with parallel sparse direct solver (PARDISO). The results obtained are presented in this section.

\subsection{Navier-Stokes}
\begin{figure}[H]
 \vspace{1pt}
 \centering
% % =================================================
% % First image, top left.
% % =================================================
 \begin{subfigure}{.5\linewidth}
     \centering
    \includegraphics[width=1.05\linewidth]{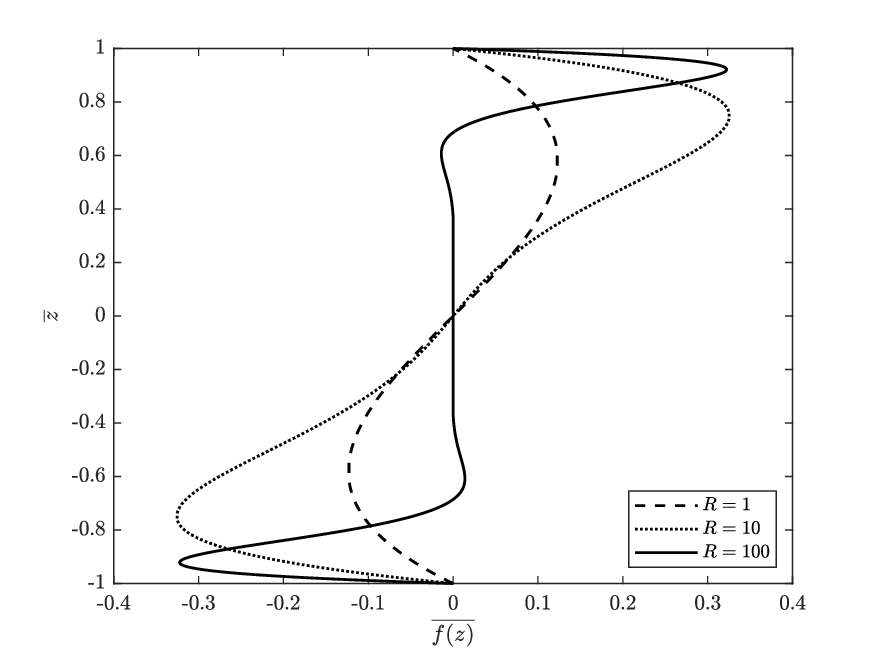}
    %\caption{$f(z)$}
    %\label{Img:1TL}
 \end{subfigure}\hfill%\\[-5ex]
% % =================================================
% % Top right
% % =================================================
 \begin{subfigure}{.5\linewidth}
    \centering
    \includegraphics[width=1.05\linewidth]{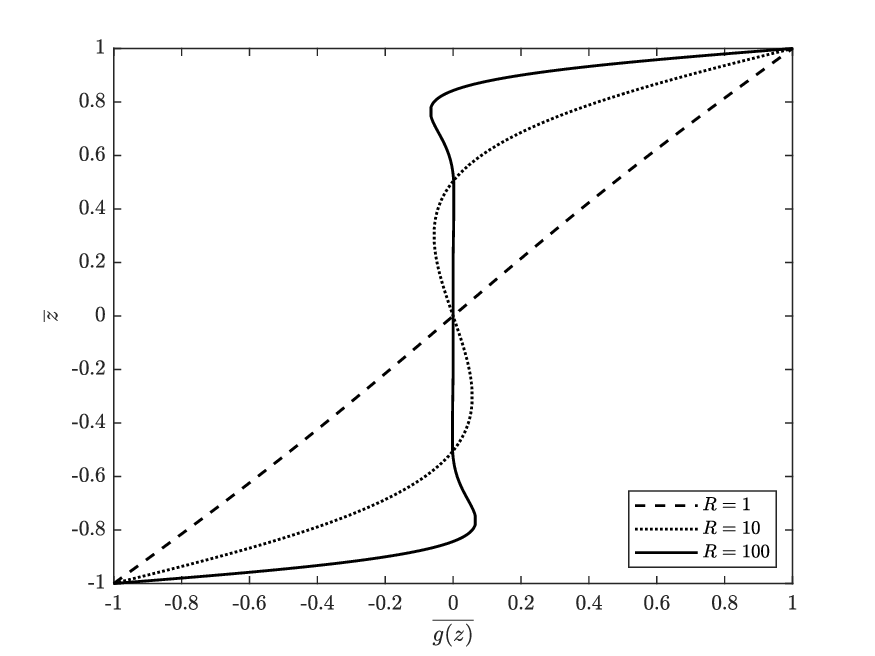}
    %\caption{$g(z)$}
    %\label{Img:1TR}
 \end{subfigure}\\[-5ex]
% % =================================================
% % Bottom left
% % =================================================
 \begin{subfigure}{.5\linewidth}
     \centering
     \includegraphics[width=1.05\linewidth]{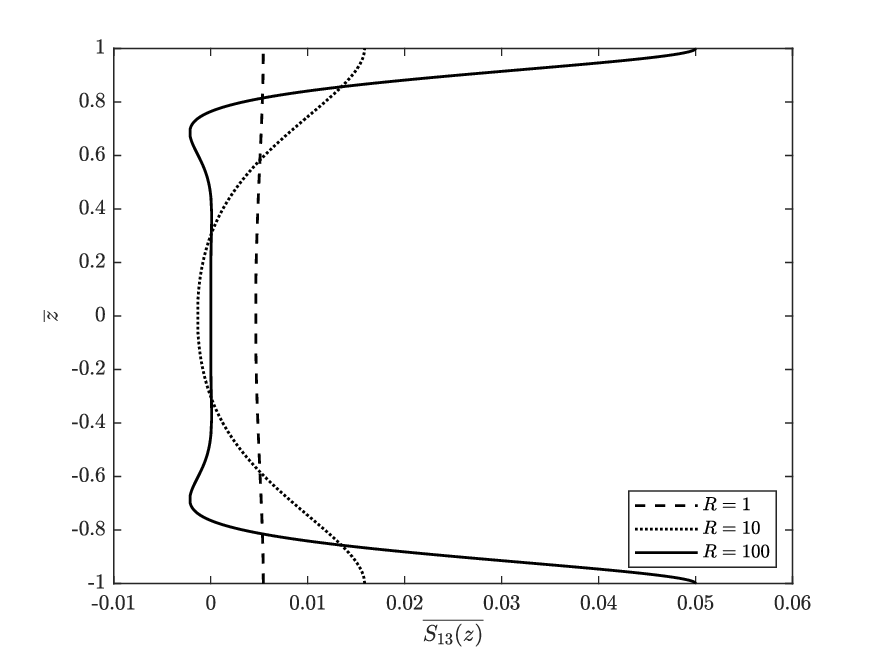}
    %\caption{$S_{13}$}
    %\label{Img:1BL}
 \end{subfigure}%\\[-5ex]
% % =================================================
% % Bottom right
% % =================================================
 \begin{subfigure}{.5\linewidth}
     \centering
     \includegraphics[width=1.05\linewidth]{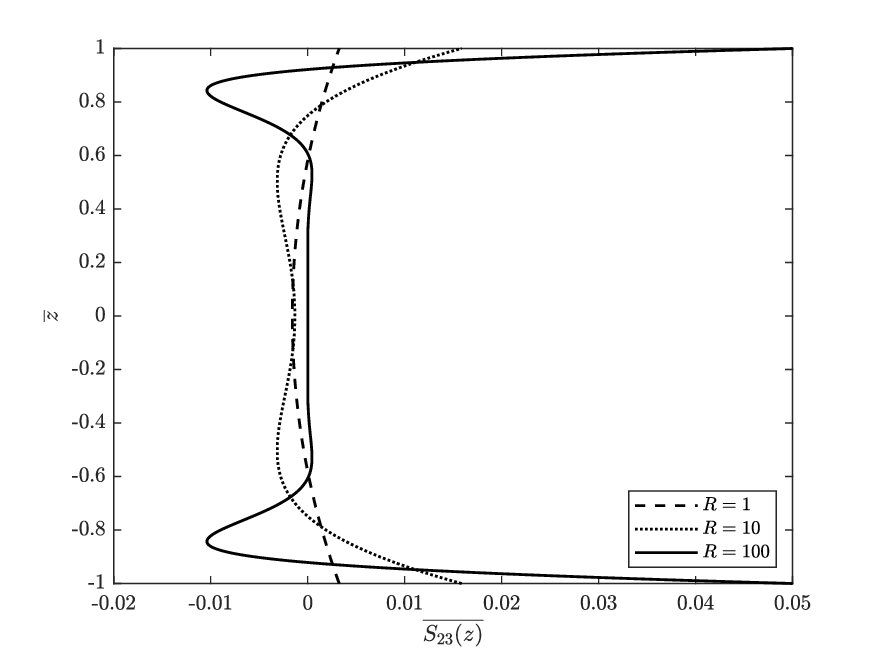}
    %\caption{$S_{23}$}
     %\label{Img:1BR}
 \end{subfigure}%\\[-5ex]
% % =================================================
% % Caption for entire figure
% % =================================================
 \caption{Solution for a Navier-Stokes Fluid at various Reynolds numbers}
 \label{Fig5.1}
 \vspace{1pt}
% % =================================================
\end{figure}

To provide a baseline for the observations discussed in the subsequent results, the solution for the Navier-Stokes fluid at various Reynolds numbers is shown in Figure \ref{Fig5.1} for $\epsilon = 0.01$. Both the systems of equations when their exponent is 0, yield the same result. It can be observed that as the Reynolds number increases, prominent boundary layers begin to appear. 

\begin{figure}[H]
 \vspace{1pt}
 \centering
 \begin{subfigure}{.5\linewidth}
     \centering
     \includegraphics[width=1.05\linewidth]{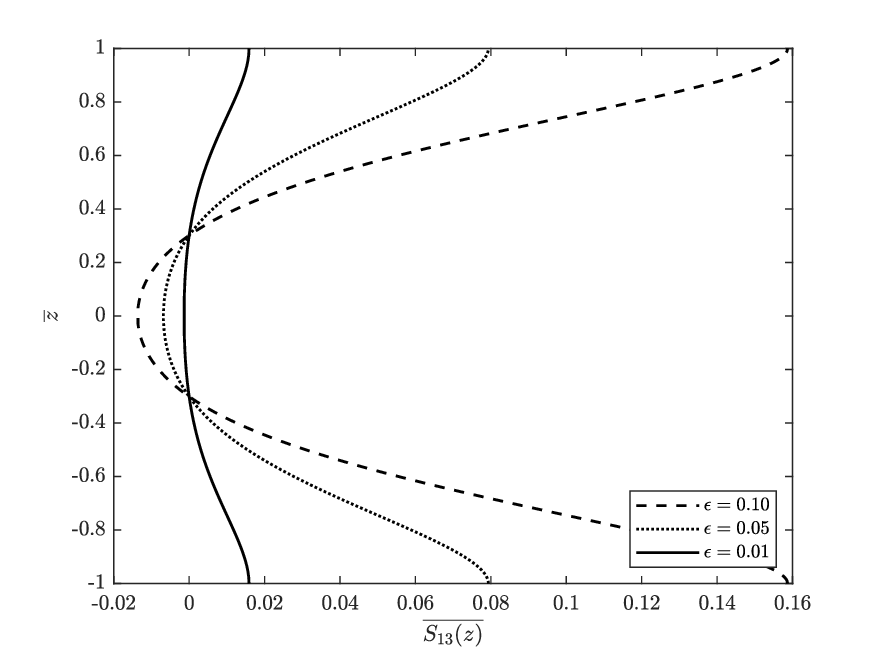}
    %\caption{$S_{13}$}
    %\label{Img:1BL}
 \end{subfigure}%\\[-5ex]
% % =================================================
% % Bottom right
% % =================================================
 \begin{subfigure}{.5\linewidth}
     \centering
     \includegraphics[width=1.05\linewidth]{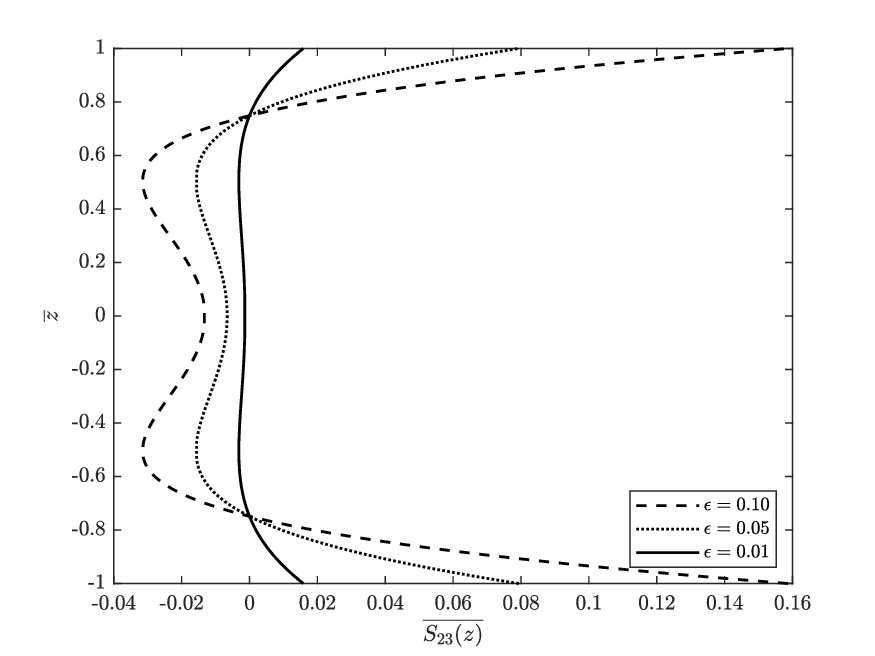}
    %\caption{$S_{23}$}
     %\label{Img:1BR}
 \end{subfigure}%\\[-5ex]
% % =================================================
% % Caption for entire figure
% % =================================================
 \caption{Stresses for a Navier-Stokes fluid for various $\epsilon$, $R = 10$}
 \label{Fig5.2}
 \vspace{1pt}
% % =================================================
\end{figure}

In order to assess the effect of the offset between the plates on the flow, we plot the solution at $R=10$ for $\epsilon = 0.01$ , $0.1$ and $1$. With increase in the offset between the plates, stress at the plates increases. This can be seen in Figure \ref{Fig5.2}.

\subsection{Stress-thinning case}
\begin{figure}[H]
 \vspace{1pt}
 \centering
% % =================================================
% % First image, top left.
% % =================================================
 \begin{subfigure}{.5\linewidth}
     \centering
    \includegraphics[width=1.05\linewidth]{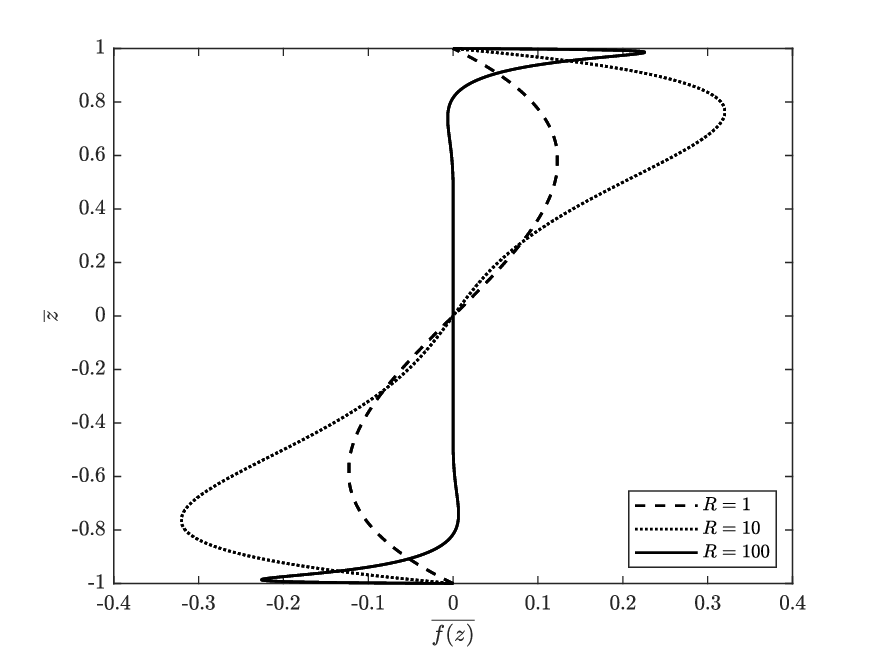}
    %\caption{$f(z)$}
    %\label{Img:1TL}
 \end{subfigure}%\\[-5ex]
% % =================================================
% % Top right
% % =================================================
 \begin{subfigure}{.5\linewidth}
    \centering
    \includegraphics[width=1.05\linewidth]{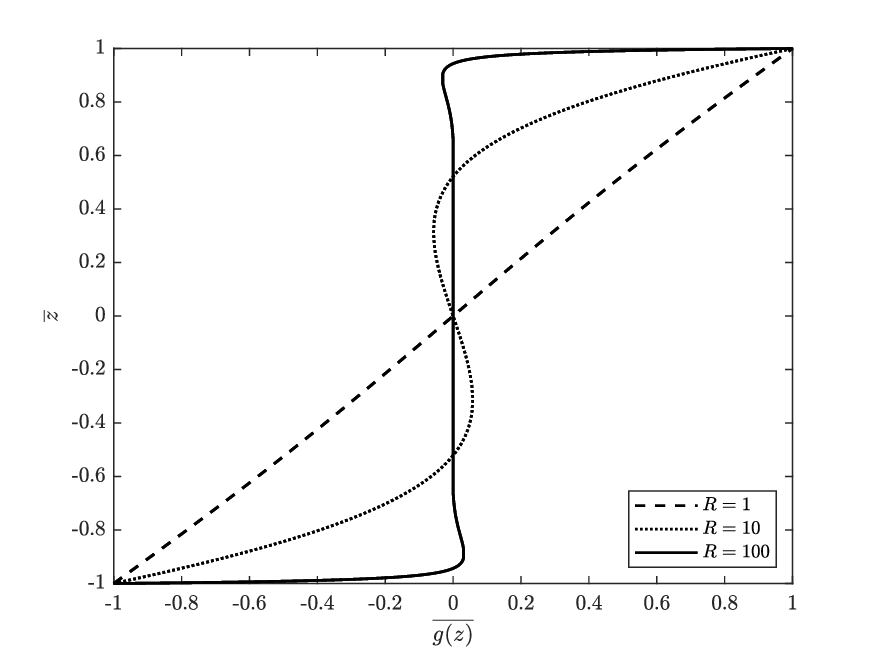}
    %\caption{$g(z)$}
    %\label{Img:1TR}
 \end{subfigure}\\%[-5ex]
% % =================================================
% % Bottom left
% % =================================================
 \begin{subfigure}{.5\linewidth}
     \centering
     \includegraphics[width=1.05\linewidth]{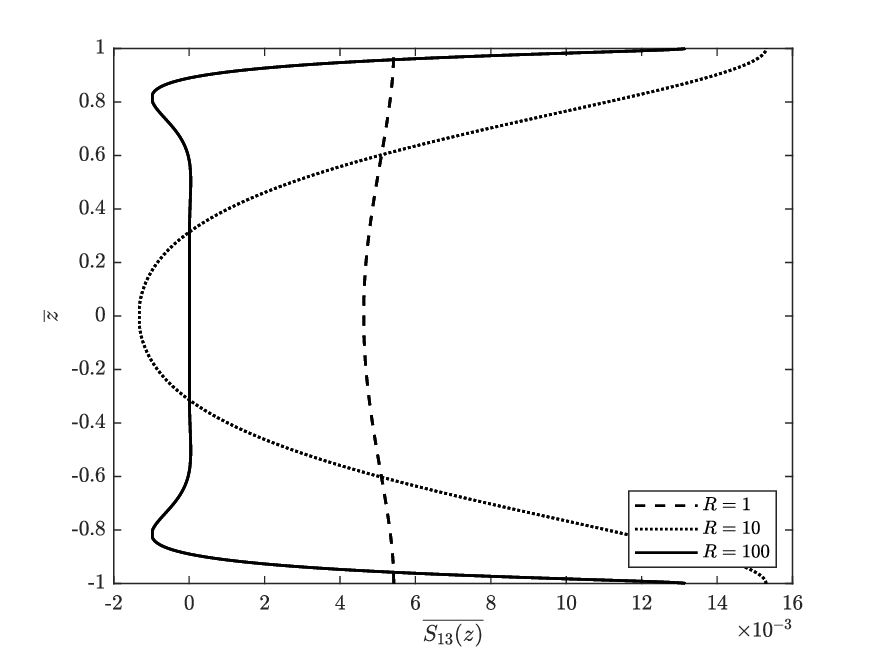}
    %\caption{$S_{13}$}
    %\label{Img:1BL}
 \end{subfigure}%\\[-5ex]
% % =================================================
% % Bottom right
% % =================================================
 \begin{subfigure}{.5\linewidth}
     \centering
     \includegraphics[width=1.05\linewidth]{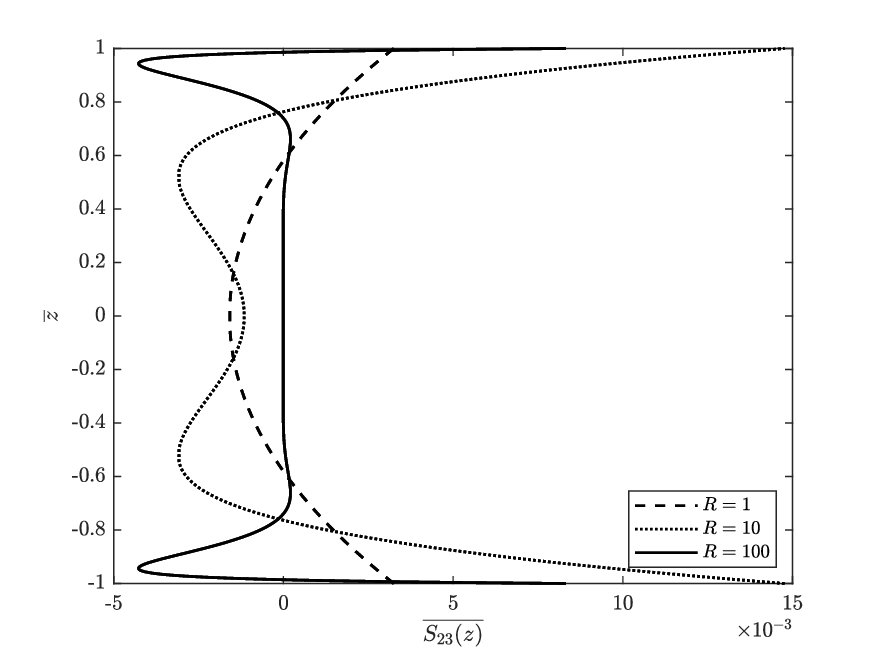}
    %\caption{$S_{23}$}
     %\label{Img:1BR}
 \end{subfigure}%\\[-5ex]
% % =================================================
% % Caption for entire figure
% % =================================================
 \caption{Solution for a stress-thinning fluid at various Reynolds numbers}
 \label{Fig5.3}
 \vspace{1pt}
% % =================================================
\end{figure}

For the generalized stress-power law fluid (\ref{Eqn1}) with $a_1 = 0$, $a_2 = 1$, and $m = 2$ we observe that the boundary layers start to appear at lower Reynolds numbers than the linearly viscous case for $\epsilon = 0.01$ (See Figure \ref{Fig5.3}). This is intuitive since increased stresses will result in an increase in generalized fluidity, resulting in higher shear rates at lower stresses. $\beta$ for this study was held constant at 1. Since $R$ is varied by varying $\Omega$, $\overline{\beta}$ varies as $R$ varies.

\begin{figure}[H]
 \vspace{1pt}
 \centering
% % =================================================
% % First image, top left.
% % =================================================
 \begin{subfigure}{.5\linewidth}
     \centering
    \includegraphics[width=1.05\linewidth]{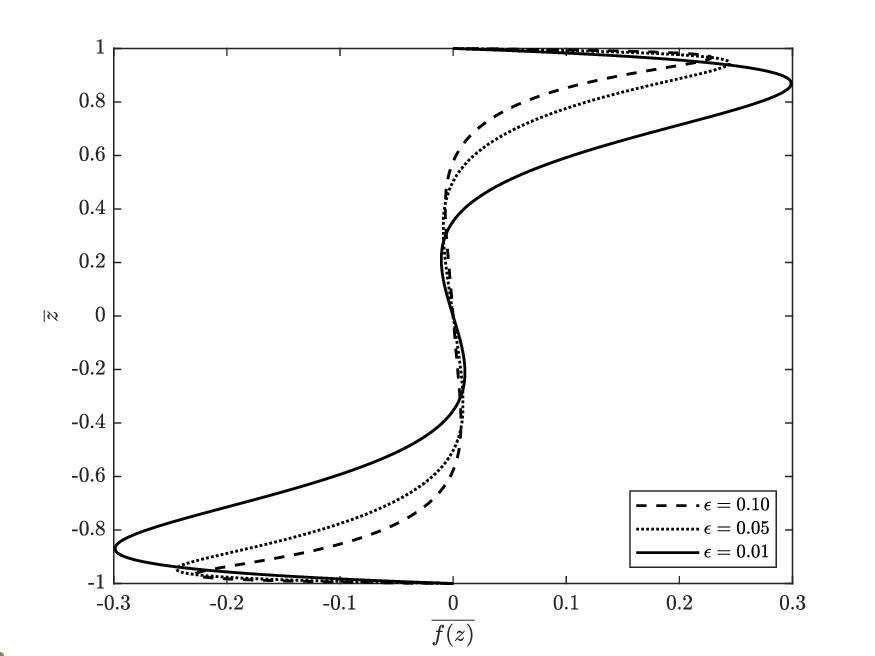}
    %\caption{$f(z)$}
    %\label{Img:1TL}
 \end{subfigure}%\\[-5ex]
% % =================================================
% % Top right
% % =================================================
 \begin{subfigure}{.5\linewidth}
    \centering
    \includegraphics[width=1.05\linewidth]{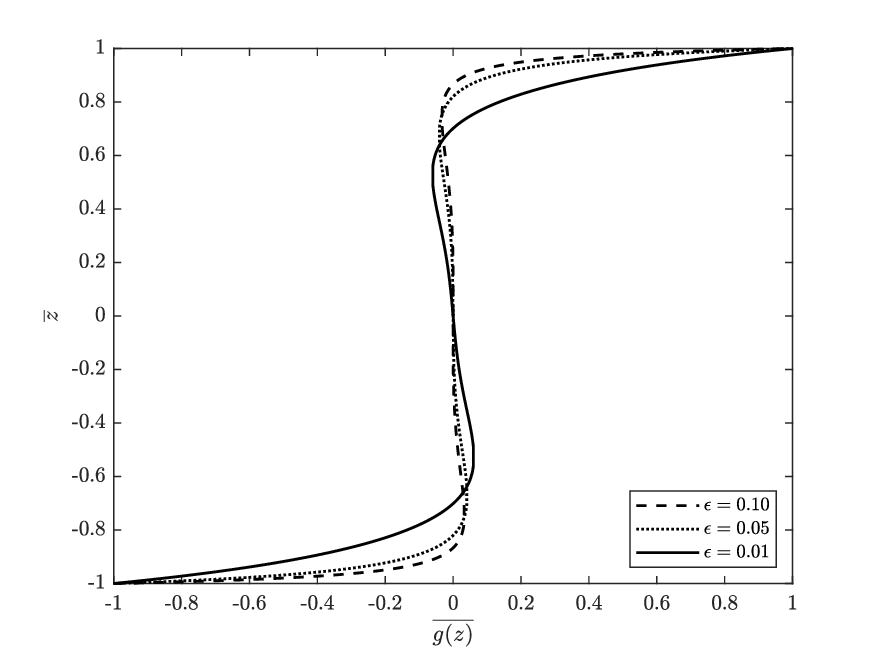}
    %\caption{$g(z)$}
    %\label{Img:1TR}
 \end{subfigure}\\ %\[ex]
% % =================================================
% % Bottom left
% % =================================================
 \begin{subfigure}{.5\linewidth}
     \centering
     \includegraphics[width=1.05\linewidth]{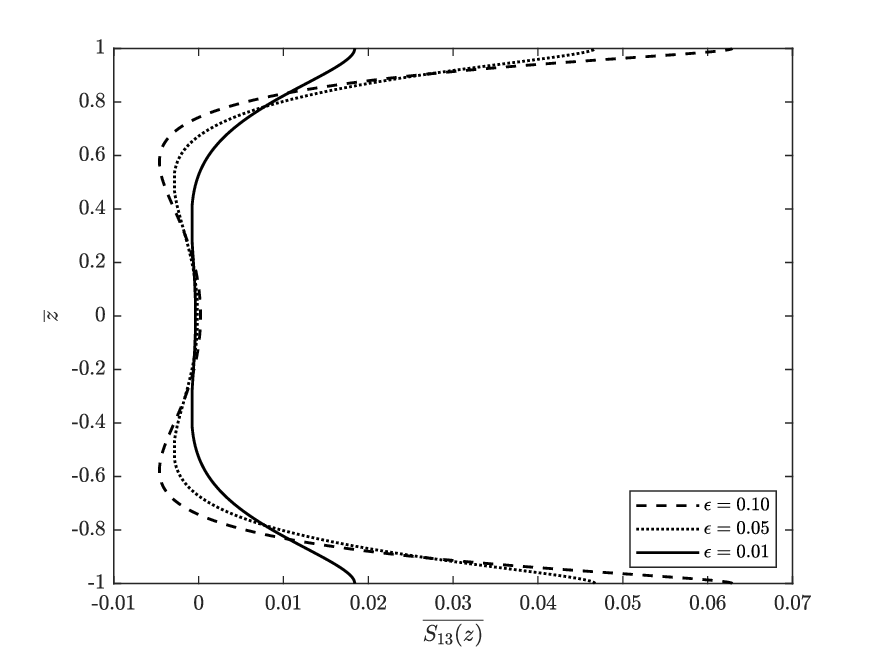}
    %\caption{$S_{13}$}
    %\label{Img:1BL}
 \end{subfigure}%\\[-5ex]
% % =================================================
% % Bottom right
% % =================================================
 \begin{subfigure}{.5\linewidth}
     \centering
     \includegraphics[width=1.05\linewidth]{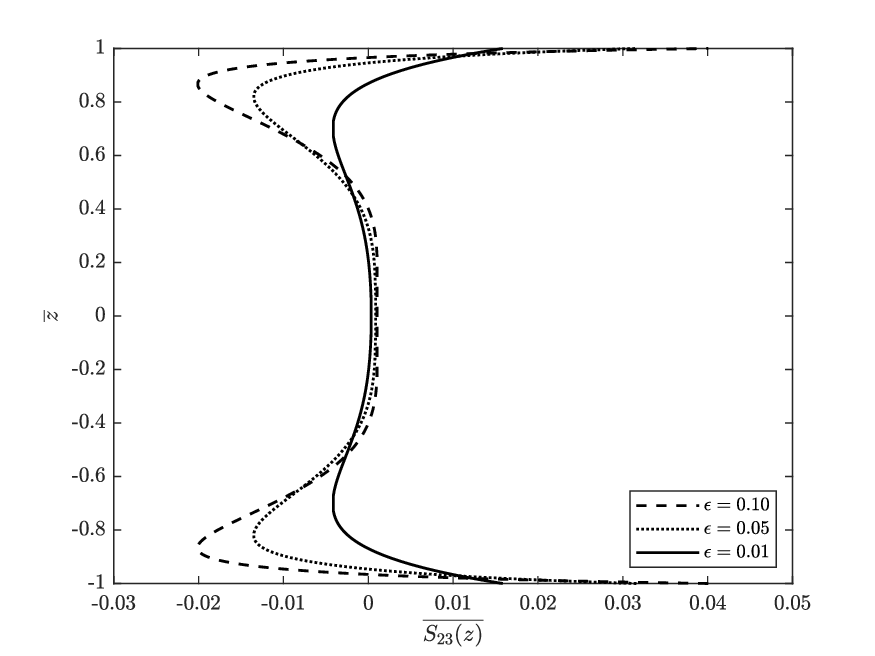}
    %\caption{$S_{23}$}
     %\label{Img:1BR}
 \end{subfigure}%\\[-5ex]
% % =================================================
% % Caption for entire figure
% % =================================================
 \caption{Solution for a stress-thinning fluid for various $\epsilon$, $R = 20$}
 \label{Fig5.4}
 \vspace{1pt}
% % =================================================
\end{figure}

Consistent with our observations in the case of linearly viscous fluid, here too we see an increase in stress as the offset between the plates increases (See Figure \ref{Fig5.4}). But this however will result in sharper boundary layers at higher $\epsilon$ for the same $R$, since the fluidity increases with increase in stress (stress-thinning).

\subsection{Stress-thickening case}

\begin{figure}[H]
 \vspace{1pt}
 \centering
% % =================================================
% % First image, top left.
% % =================================================
 \begin{subfigure}{.5\linewidth}
     \centering
    \includegraphics[width=1.05\linewidth]{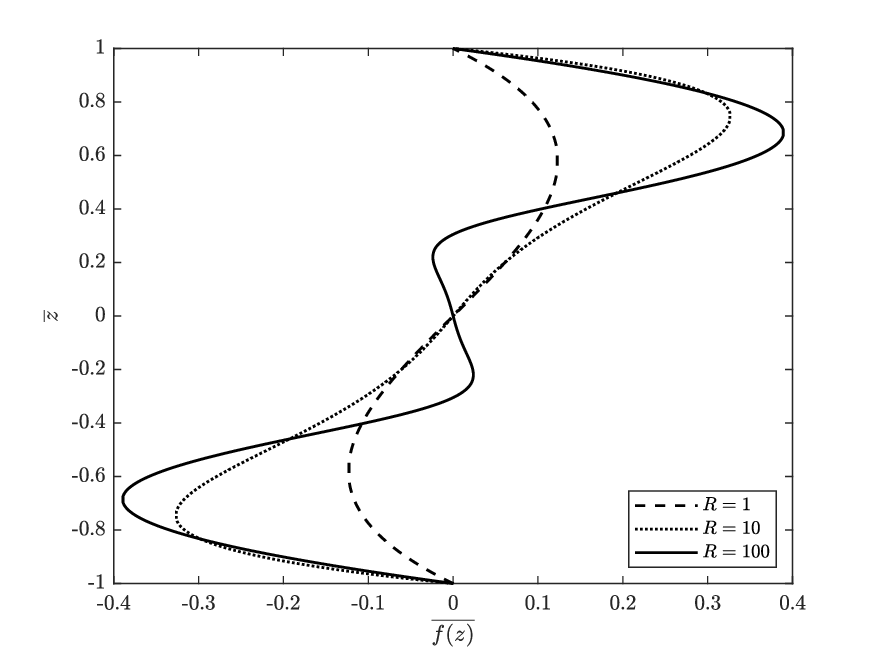}
    %\caption{$f(z)$}
    %\label{Img:1TL}
 \end{subfigure}%\\[-5ex]
% % =================================================
% % Top right
% % =================================================
 \begin{subfigure}{.5\linewidth}
    \centering
    \includegraphics[width=1.05\linewidth]{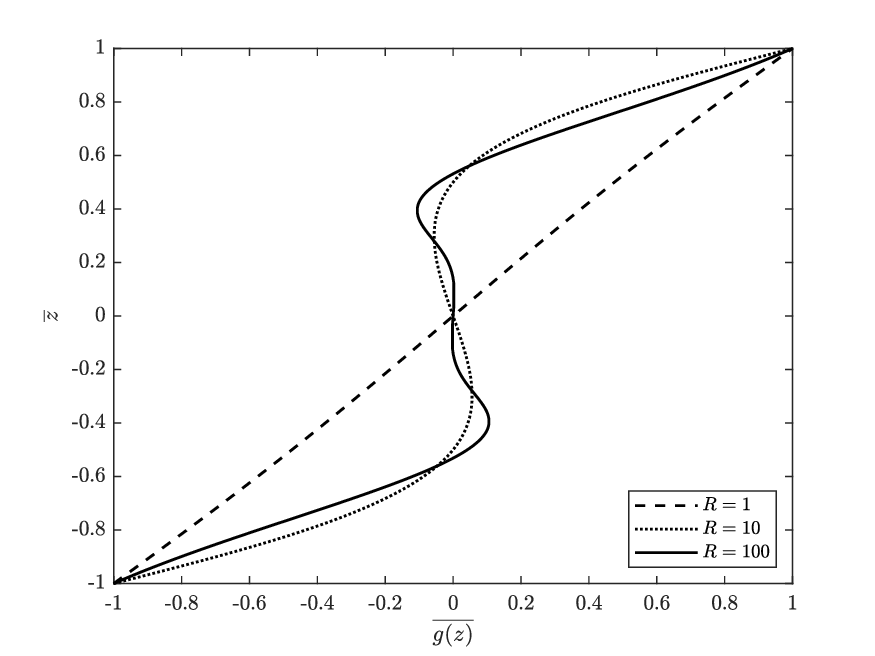}
    %\caption{$g(z)$}
    %\label{Img:1TR}
 \end{subfigure}\\[-5ex]
% % =================================================
% % Bottom left
% % =================================================
 \begin{subfigure}{.5\linewidth}
     \centering
     \includegraphics[width=1.05\linewidth]{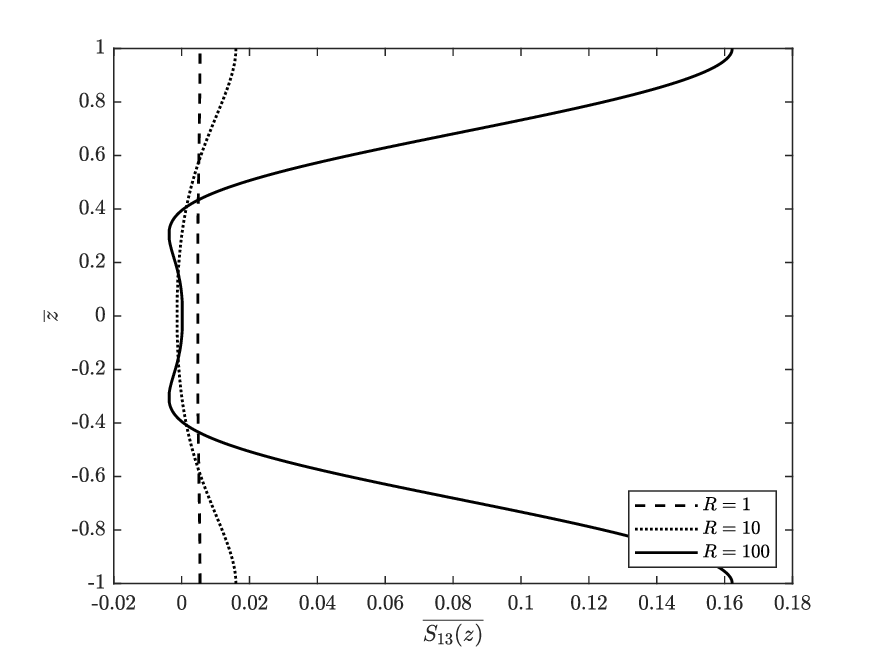}
    %\caption{$S_{13}$}
    %\label{Img:1BL}
 \end{subfigure}%\\[-5ex]
% % =================================================
% % Bottom right
% % =================================================
 \begin{subfigure}{.5\linewidth}
     \centering
     \includegraphics[width=1.05\linewidth]{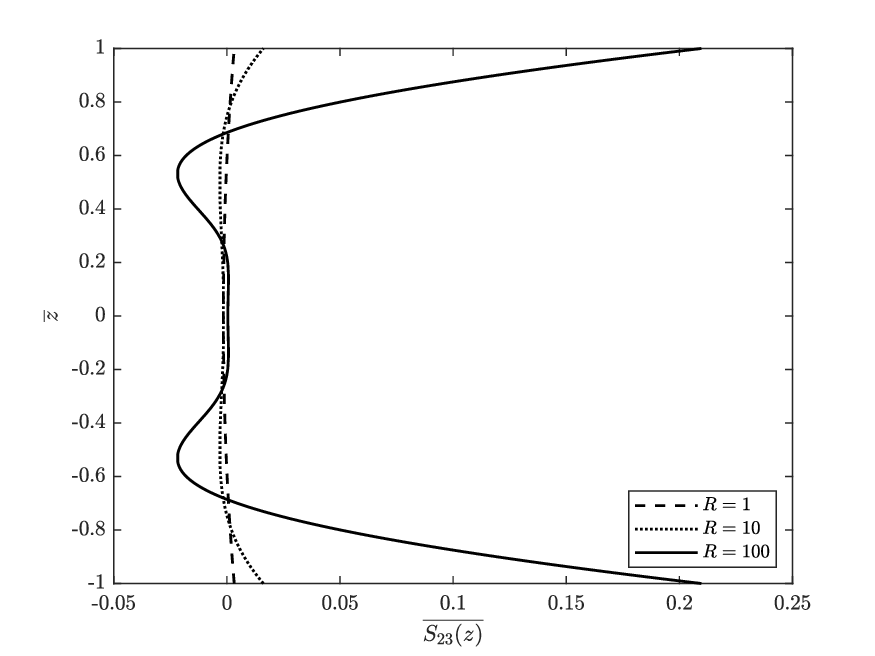}
    %\caption{$S_{23}$}
     %\label{Img:1BR}
 \end{subfigure}%\\[-5ex]
% % =================================================
% % Caption for entire figure
% % =================================================
 \caption{Solution for a stress-thickening fluid at various Reynolds numbers}
 \label{Fig5.5}
 \vspace{1pt}
% % =================================================
\end{figure}

When $a_1 = 0$, $a_2 = 1$, and $m = -0.4$, model (\ref{Eqn1}) exhibits stress-thickening behaviour. It can be seen that much higher stresses are recorded than in the stress-thinning case for the same $R$ for $\epsilon = 0.01$ (See Figure \ref{Fig5.5}). This is because the fluidity of the material reduces as the stresses increase. Here, $\|\mathbb{D}\|$ does not become zero even at $R = 100$ indicating the absence of a boundary layer. $\beta$ was held constant at 1 for this case too.

\begin{figure}[H]
 \vspace{1pt}
 \centering
% % =================================================
% % First image, top left.
% % =================================================
 \begin{subfigure}{.5\linewidth}
     \centering
    \includegraphics[width=1.05\linewidth]{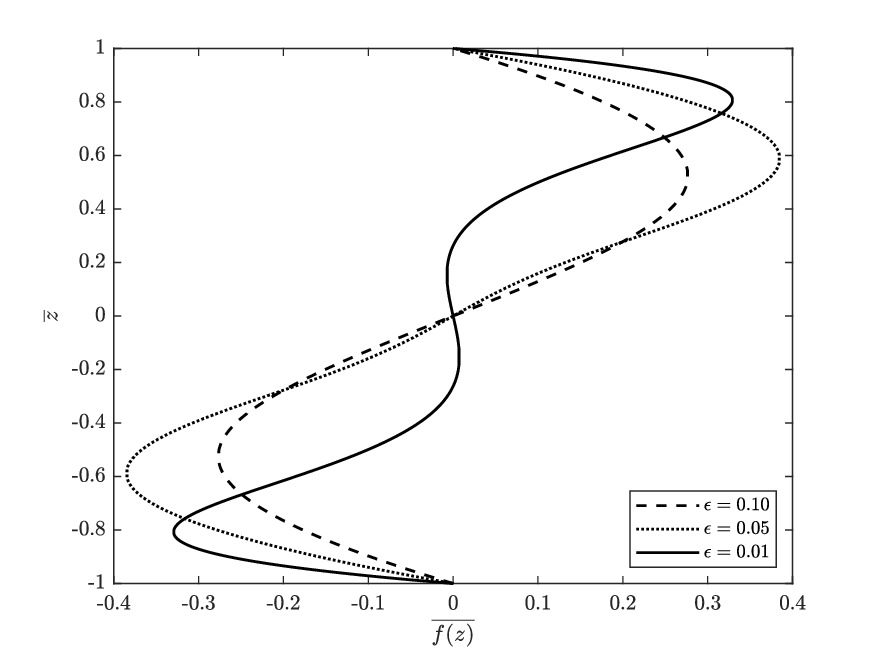}
    %\caption{$f(z)$}
    %\label{Img:1TL}
 \end{subfigure}%\\[-5ex]
% % =================================================
% % Top right
% % =================================================
 \begin{subfigure}{.5\linewidth}
    \centering
    \includegraphics[width=1.05\linewidth]{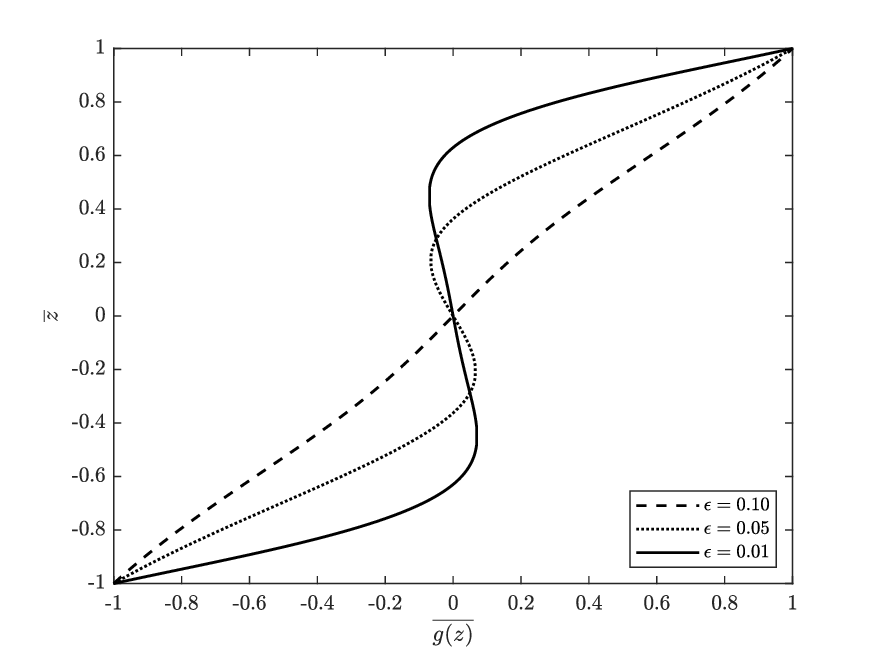}
    %\caption{$g(z)$}
    %\label{Img:1TR}
 \end{subfigure}\\[-5ex]
% % =================================================
% % Bottom left
% % ================================================
 \begin{subfigure}{.5\linewidth}
     \centering
     \includegraphics[width=1.05\linewidth]{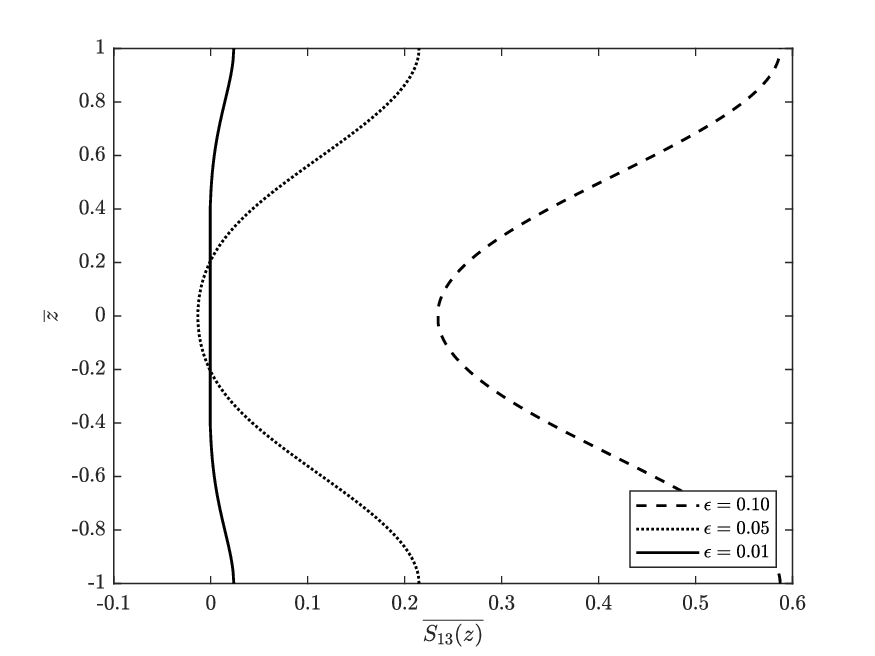}
    %\caption{$S_{13}$}
    %\label{Img:1BL}
 \end{subfigure}%\\[-5ex]
% % =================================================
% % Bottom right
% % =================================================
 \begin{subfigure}{.5\linewidth}
     \centering
     \includegraphics[width=1.05\linewidth]{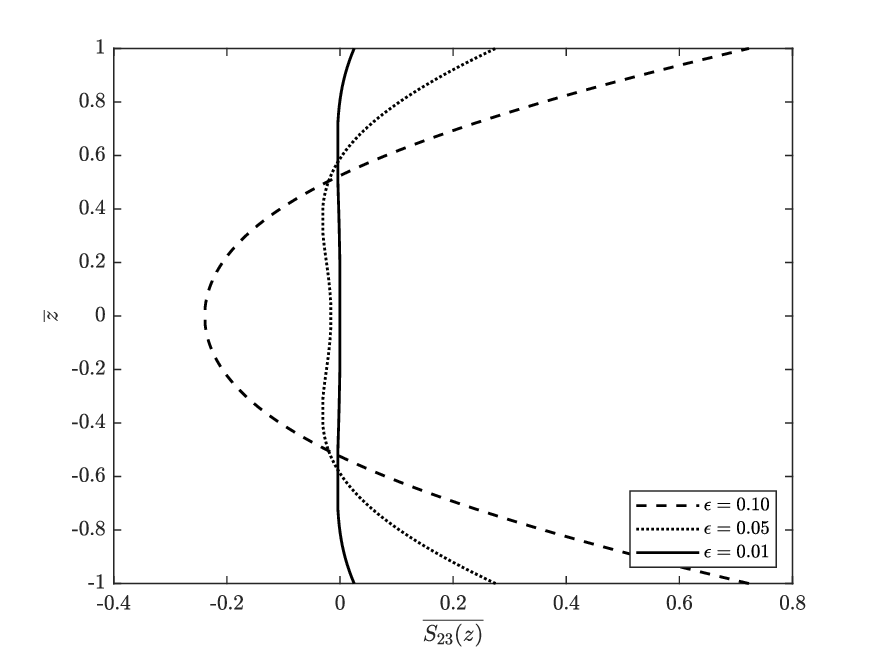}
    %\caption{$S_{23}$}
     %\label{Img:1BR}
 \end{subfigure}%\\[-5ex]
% % =================================================
% % Caption for entire figure
% % =================================================
 \caption{Solution for a stress-thickening fluid for various $\epsilon$, $R = 20$}
 \label{Fig5.6}
 \vspace{1pt}
% % =================================================
\end{figure}

Because of the stress-thickening behaviour, $\overline{f}$ and $\overline{g}$ show steeper slopes with respect to $\overline{z}$ near the plates as the offset between the plates decreases (See Figure \ref{Fig5.6}). This is essentially what happens even if we increase the Reynolds number for a fixed offset. Due to inertial effects, the slope of $\overline{f}$ and $\overline{g}$ with respect to $\overline{z}$ near the plate initially increases until a critical $R$ but then again starts to decrease.

\subsection{Non-monotone constitutive relation with one inflection point}

\begin{figure}[H]
 \vspace{1pt}
 \centering
% % =================================================
% % First image, top left.
% % =================================================
 \begin{subfigure}{.5\linewidth}
     \centering
    \includegraphics[width=1.05\linewidth]{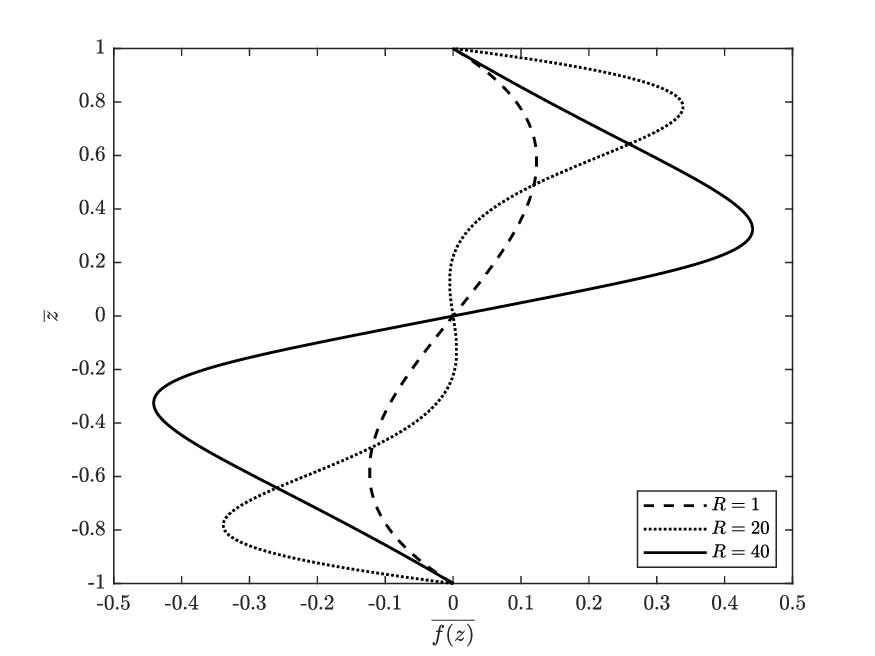}
    %\caption{$f(z)$}
    %\label{Img:1TL}
 \end{subfigure}%\\[-5ex]
% % =================================================
% % Top right
% % =================================================
 \begin{subfigure}{.5\linewidth}
    \centering
    \includegraphics[width=1.05\linewidth]{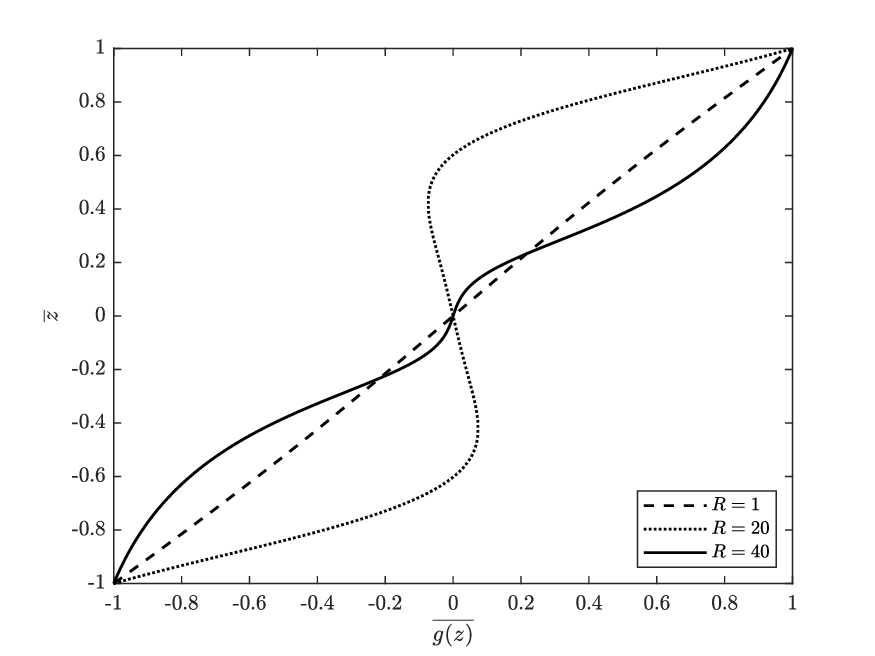}
    %\caption{$g(z)$}
    %\label{Img:1TR}
 \end{subfigure}\\[-5ex]
% % =================================================
% % Bottom left
% % =================================================
 \begin{subfigure}{.5\linewidth}
     \centering
     \includegraphics[width=1.05\linewidth]{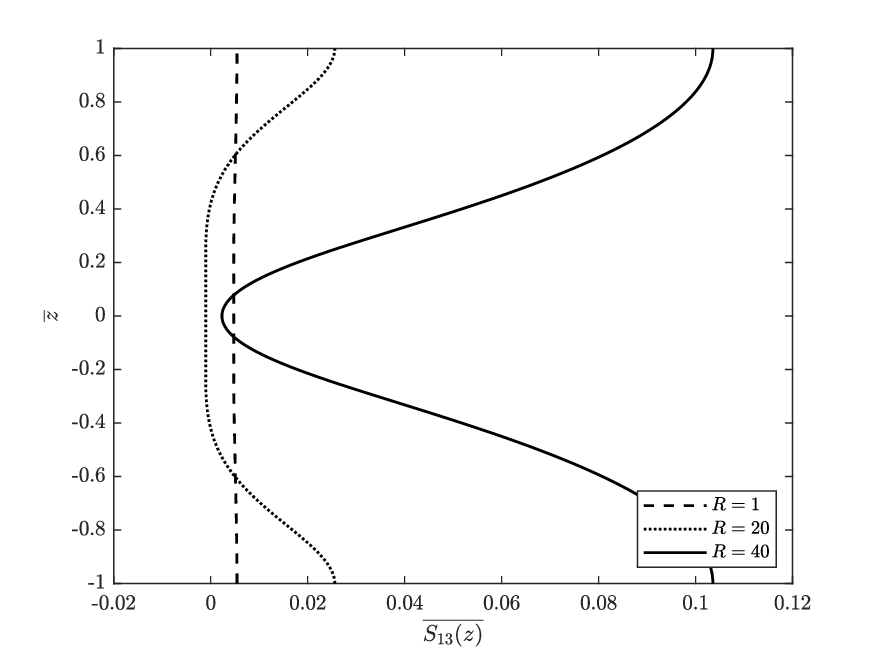}
    %\caption{$S_{13}$}
    %\label{Img:1BL}
 \end{subfigure}%\\[-5ex]
% % =================================================
% % Bottom right
% % =================================================
 \begin{subfigure}{.5\linewidth}
     \centering
     \includegraphics[width=1.05\linewidth]{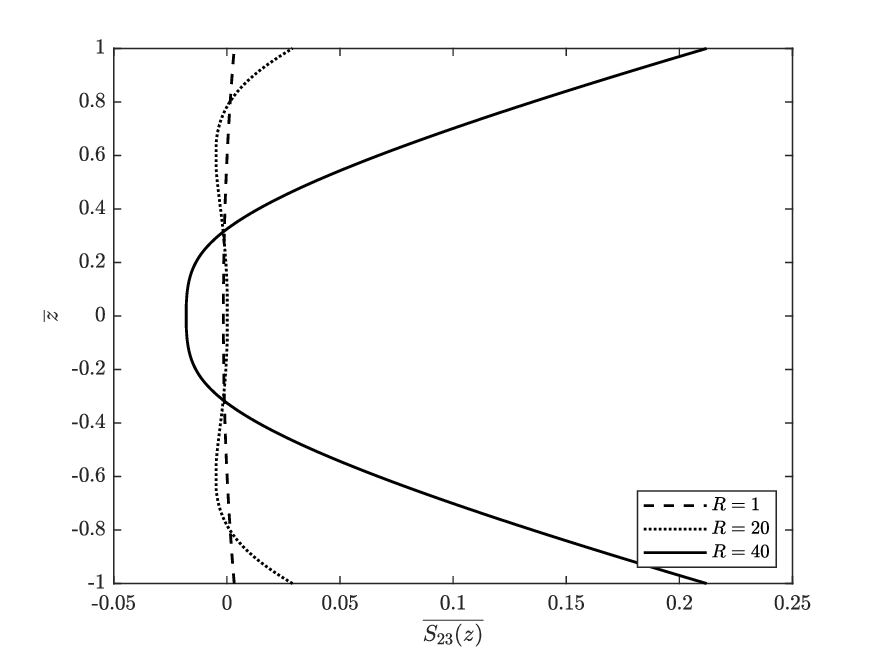}
    %\caption{$S_{23}$}
     %\label{Img:1BR}
 \end{subfigure}%\\[-5ex]
% % =================================================
% % Caption for entire figure
% % =================================================
 \caption{Solution for a stress power-law fluid with a non-monotonic relation having one inflection point at various Reynolds numbers for $m = -0.8$}
 \label{Fig5.7}
 \vspace{1pt}
% % =================================================
\end{figure}

\begin{figure}[H]
 \vspace{1pt}
 \centering
% % =================================================
% % First image, top left.
% % =================================================
 \begin{subfigure}{.5\linewidth}
     \centering
    \includegraphics[width=1.05\linewidth]{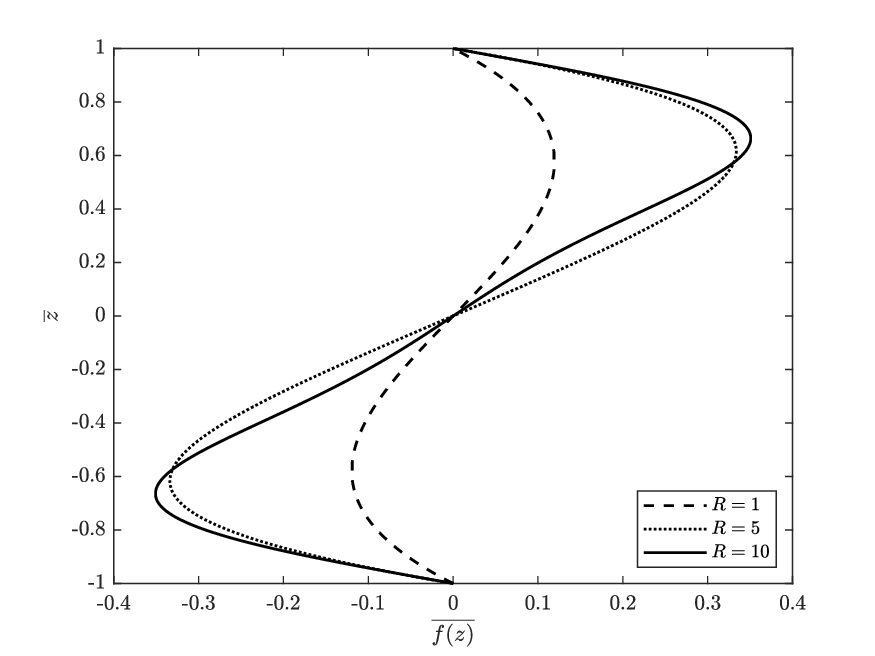}
    %\caption{$f(z)$}
    %\label{Img:1TL}
 \end{subfigure}%\\[-5ex]
% % =================================================
% % Top right
% % =================================================
 \begin{subfigure}{.5\linewidth}
    \centering
    \includegraphics[width=1.05\linewidth]{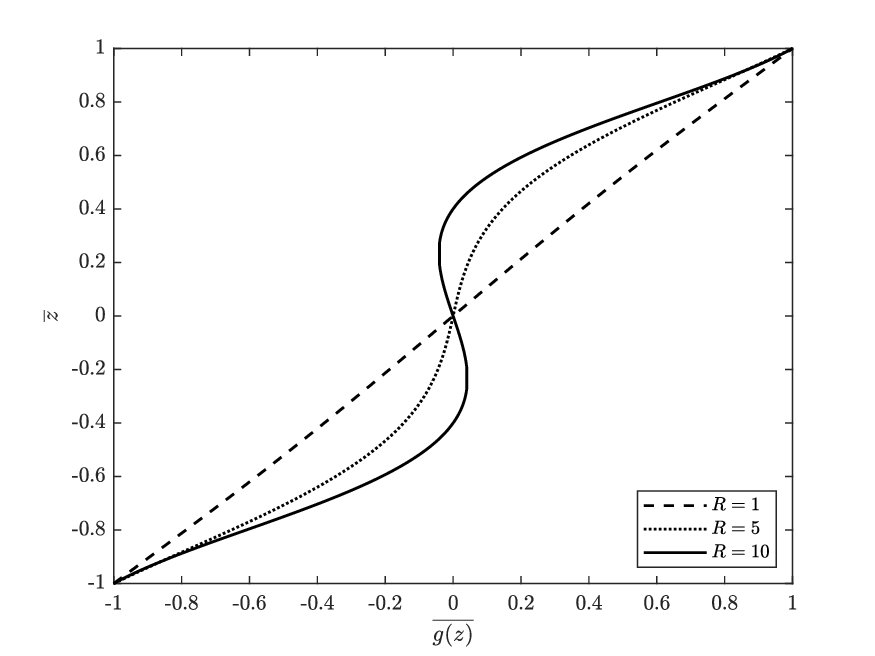}
    %\caption{$g(z)$}
    %\label{Img:1TR}
 \end{subfigure}\\[-5ex]
% % =================================================
% % Bottom left
% % =================================================
 \begin{subfigure}{.5\linewidth}
     \centering
     \includegraphics[width=1.05\linewidth]{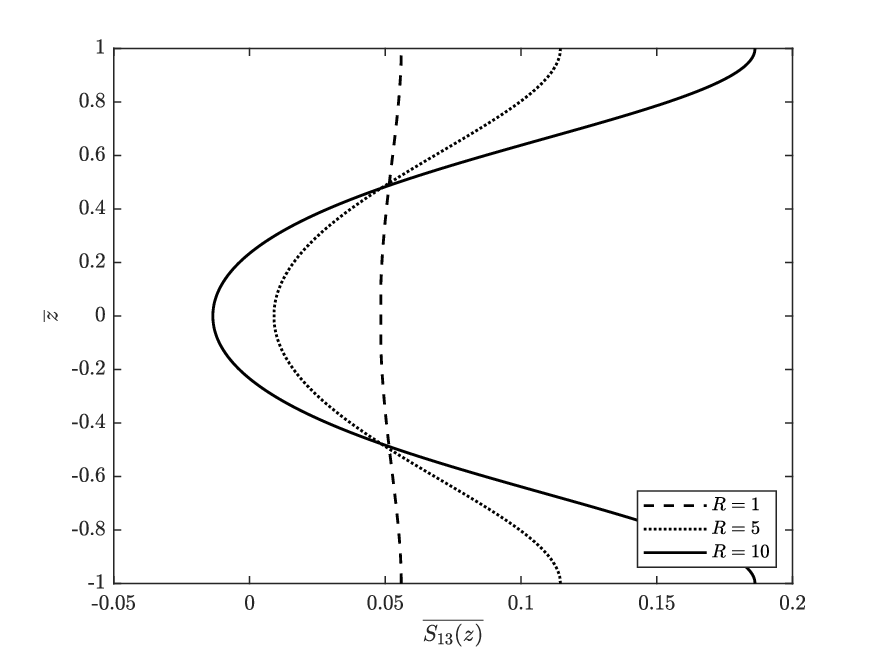}
    %\caption{$S_{13}$}
    %\label{Img:1BL}
 \end{subfigure}%\\[-5ex]
% % =================================================
% % Bottom right
% % =================================================
 \begin{subfigure}{.5\linewidth}
     \centering
     \includegraphics[width=1.05\linewidth]{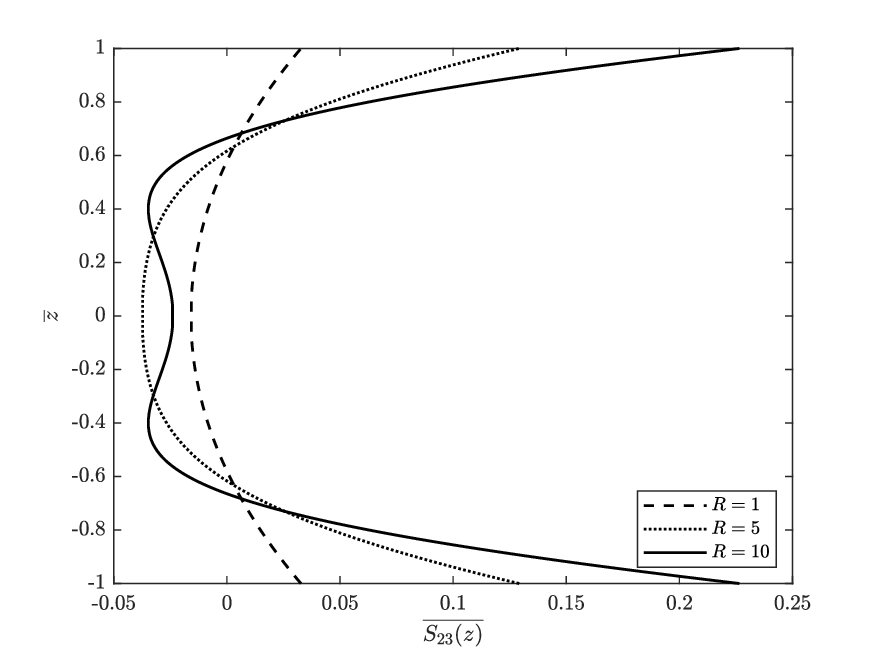}
    %\caption{$S_{23}$}
     %\label{Img:1BR}
 \end{subfigure}%\\[-5ex]
% % =================================================
% % Caption for entire figure
% % =================================================
 \caption{Solution for a stress power-law fluid when $\overline{\beta}$ is held constant instead of $\beta$}
 \label{Fig31}
 \vspace{1pt}
% % =================================================
\end{figure}

When $a_1 = 0$, $a_2 = 1$, and $m<-0.5$ in (\ref{Eqn1}), the model exhibits a non-monotonic relationship between $\|\mathbb{D}\|$ and $\|\mathbb{S}\|$. Strain rate increases with increasing stress until a critical stress after which it hits an inflection point, for there the strain rate starts to reduce with increasing stress asymptotically reaching zero. Changes in the flow as the Reynolds number increases has been shown in Figure \ref{Fig5.7}. $\beta$ and $\epsilon$ have been held constant at 1 and 0.01 respectively. Although this model has been studied by Srinivasan and Karra \cite{srinivasan2015flow}, there $\overline{\beta}$ has been held constant instead of $\beta$. Which means with increasing $R$, $\beta$ which is a material parameter has to vary to maintain a constant $\overline{\beta}$ while accounting for increasing $\Omega$. This essentially means different fluids flowing at different Reynolds numbers are being compared. Since the fluid under consideration is fixed, it is more appropriate to consider the case when $\beta$ is held constant rather than when $\overline{\beta}$ is held constant while we vary  $R$. Nevertheless, we were able to obtain the same result as that of Figure 3 in \cite{srinivasan2015flow} when we set $\epsilon = 0.1$, $a_1 = 0$, $a_2 = 1$, $m = -6$, and $\overline{\beta}=1$. (See Figure \ref{Fig31})

\begin{figure}[H]
    \centering
    \includegraphics[scale = 0.9]{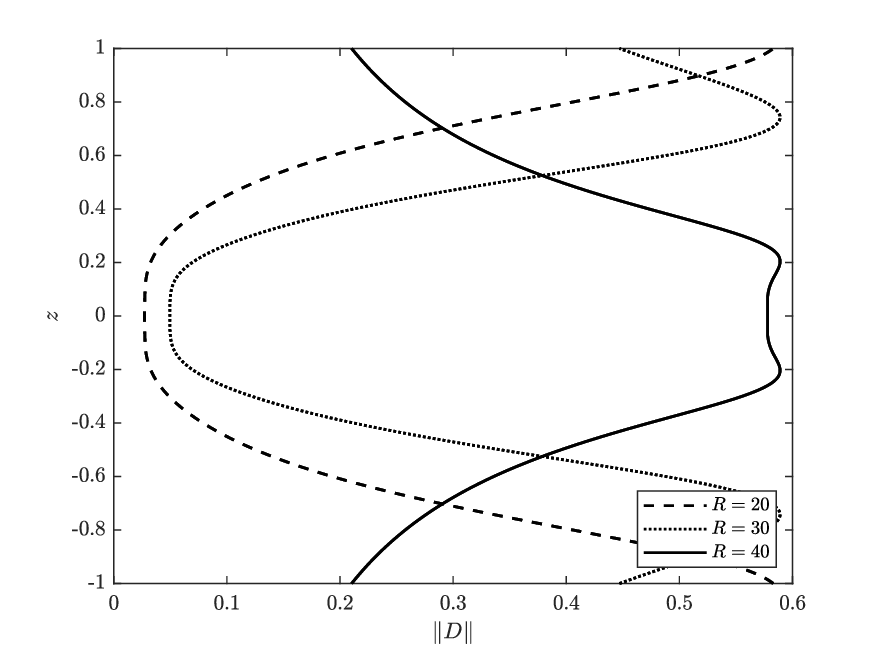}
    \caption{$\|\mathbb{D}\|$ as $R$ varies when $a_1 = 0$, $m = -0.8$}
    \label{Fig5.8}
\end{figure}

As we increase the Reynolds number, we observe that the region with maximum $\|\mathbb{D}\|$ is no longer at the top and bottom plates but it moves towards the center of the domain after a critical value of $R$. This is immediately clear from the $\|\mathbb{D}\|$ vs $z$ plot shown in Figure \ref{Fig5.8}.

\subsection{S-type non-monotone constitutive relation}

\begin{figure}[H]
 \vspace{1pt}
 \centering
% % =================================================
% % First image, top left.
% % =================================================
 \begin{subfigure}{.5\linewidth}
     \centering
    \includegraphics[width=1.05\linewidth]{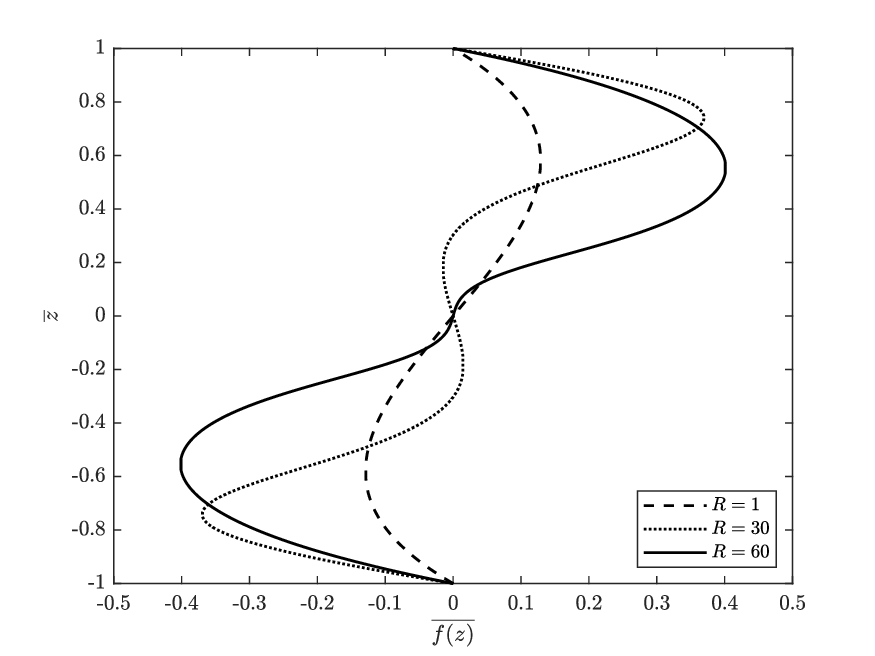}
    %\caption{$f(z)$}
    %\label{Img:1TL}
 \end{subfigure}%\\[-5ex]
% % =================================================
% % Top right
% % =================================================
 \begin{subfigure}{.5\linewidth}
    \centering
    \includegraphics[width=1.05\linewidth]{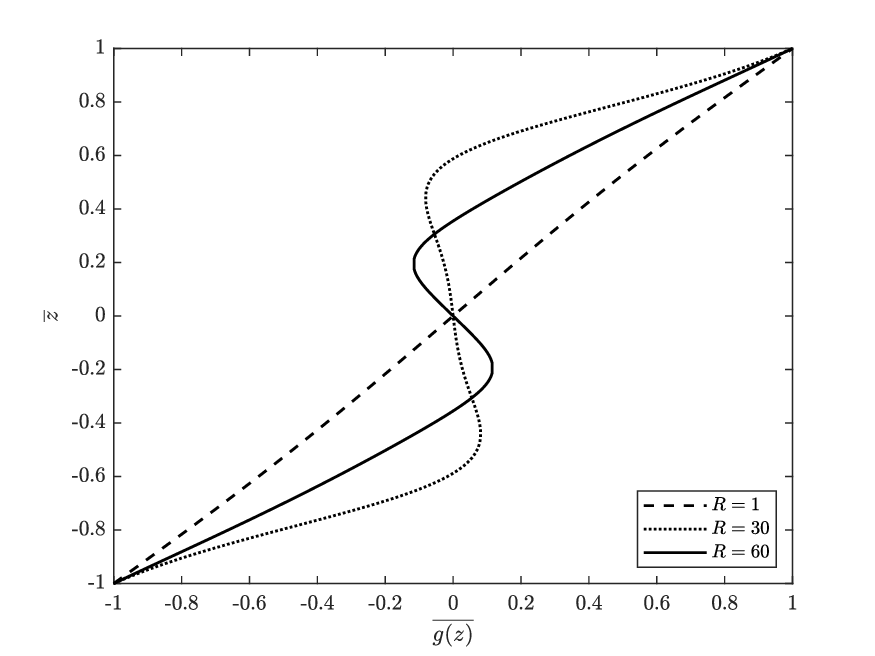}
    %\caption{$g(z)$}
    %\label{Img:1TR}
 \end{subfigure}\\[-5ex]
% % =================================================
% % Bottom left
% % =================================================
 \begin{subfigure}{.5\linewidth}
     \centering
     \includegraphics[width=1.05\linewidth]{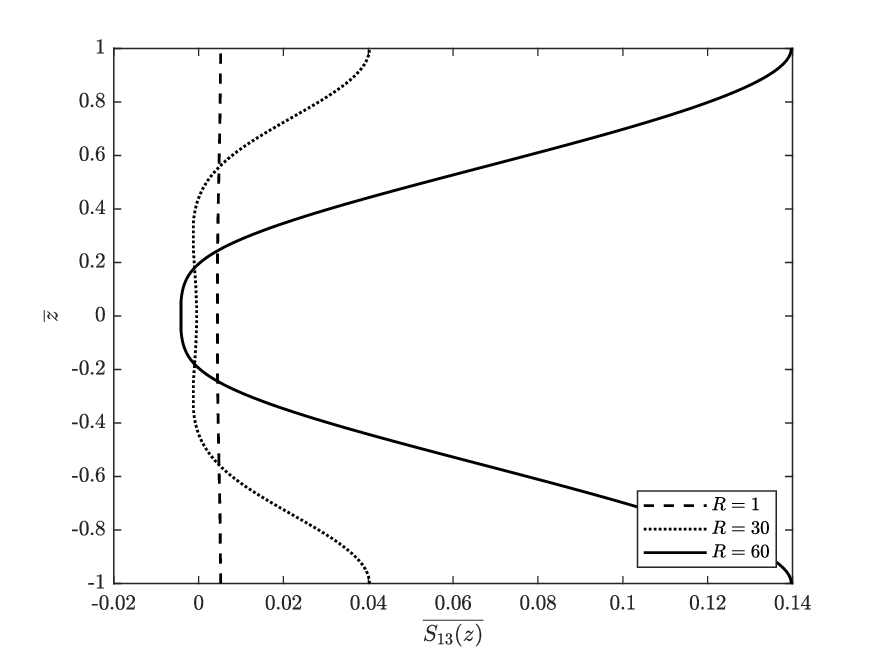}
    %\caption{$S_{13}$}
    %\label{Img:1BL}
 \end{subfigure}%\\[-5ex]
% % =================================================
% % Bottom right
% % =================================================
 \begin{subfigure}{.5\linewidth}
     \centering
     \includegraphics[width=1.05\linewidth]{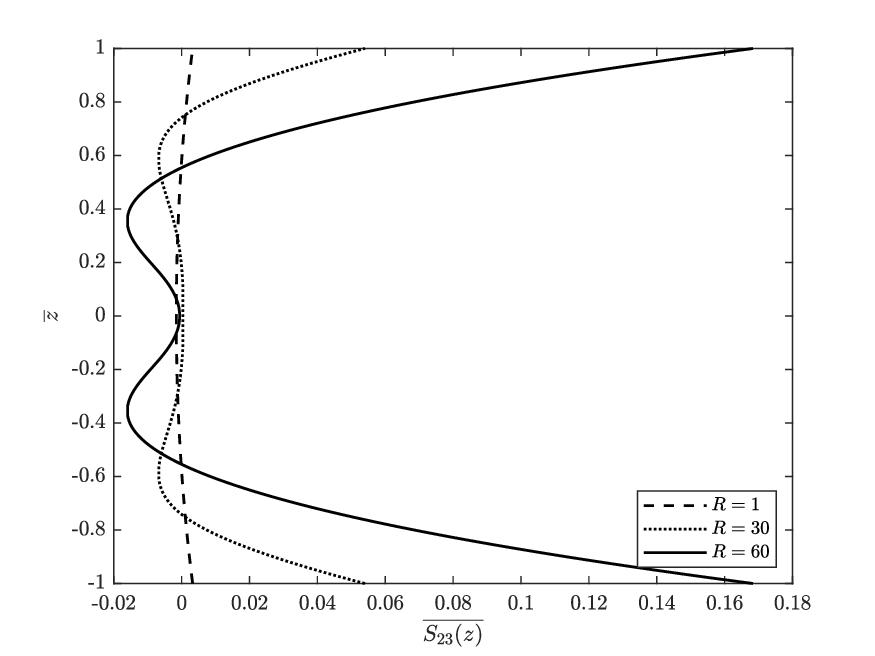}
    %\caption{$S_{23}$}
     %\label{Img:1BR}
 \end{subfigure}%\\[-5ex]
% % =================================================
% % Caption for entire figure
% % =================================================
 \caption{Solution for a stress power-law fluid with a non-monotonic relation having two inflection point at various Reynolds numbers for $m = -0.8$ and $a_1 = 0.05$}
 \label{Fig5.9}
 \vspace{1pt}
% % =================================================
\end{figure}

\begin{figure}[H]
 \vspace{1pt}
 \centering
% % =================================================
% % Bottom left
% % =================================================
 \begin{subfigure}{.5\linewidth}
     \centering
     \includegraphics[width=1.05\linewidth]{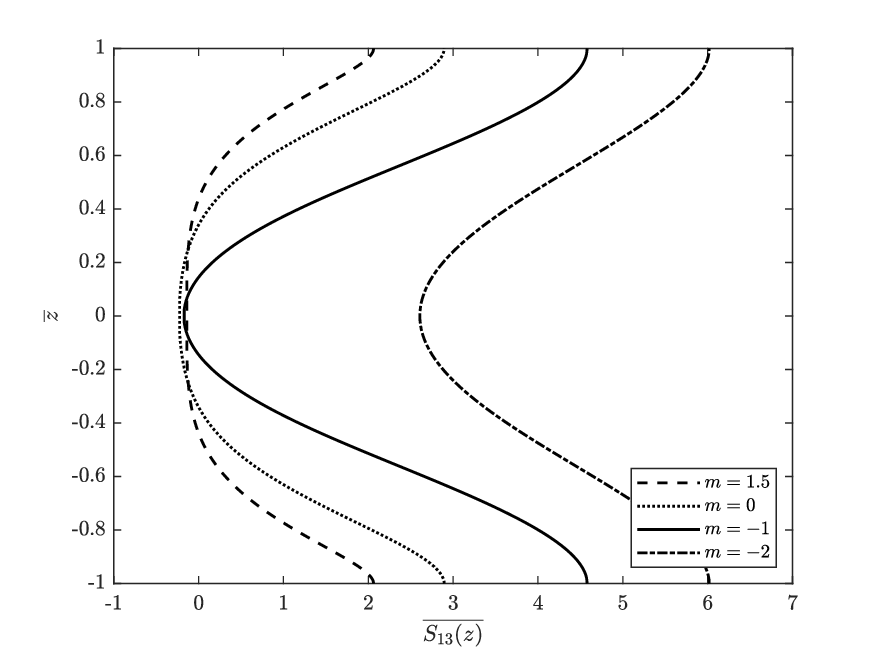}
    %\caption{$S_{13}$}
    %\label{Img:1BL}
 \end{subfigure}%\\[-5ex]
% % =================================================
% % Bottom right
% % =================================================
 \begin{subfigure}{.5\linewidth}
     \centering
     \includegraphics[width=1.05\linewidth]{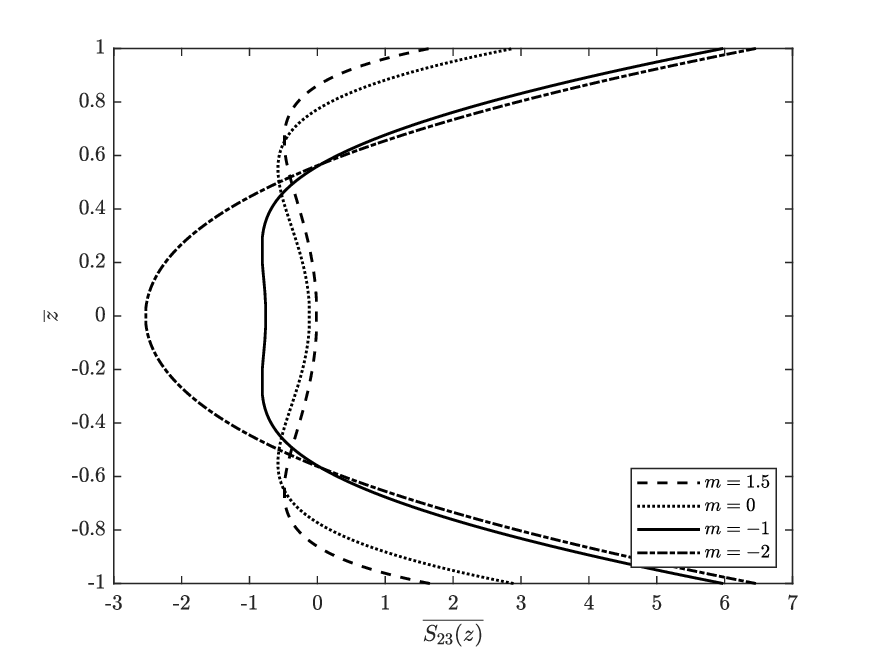}
    %\caption{$S_{23}$}
     %\label{Img:1BR}
 \end{subfigure}%\\[-5ex]
% % =================================================
% % Caption for entire figure
% % =================================================
 \caption{Dependence of stresses on the power-law exponent $m$}
 \label{Fig33}
 \vspace{1pt}
% % =================================================
\end{figure}

For $a_2 = 1$, $m<-0.5$, and $0<a_1<2\left(\frac{\left|2m+1\right|}{2\left(1-m\right)}\right)^{\left(1-m\right)}$ the model exhibits an S-type non-monotonic relationship between $\|\mathbb{D}\|$ and $\|\mathbb{S}\|$ with two inflection points. Solution for such a model as the Reynolds number increases has been shown in Figure \ref{Fig5.9}. $\beta$ and $\epsilon$ have been held constant at 1 and 0.01 respectively just like the previous studies. Recently this model was studied by Fusi et al. \cite{fusi2023flow}. However, that study too has been done by holding $\overline{\beta}$ constant instead of $\beta$. In \cite{fusi2023flow}, the authors have remarked that in regions with three different stresses resulting in same strain-rate, irrespective of the initial guess, the solution always converged to the least stress value deeming the other two solutions unstable akin to the strain-rate controlled experiments reported by Boltenhagen et al. \cite{boltenhagen1997observation}. We believe that this might be the case because $\overline{\beta}$ has been held constant. Because it means that the angular velocity $\Omega$ was held constant and the material parameters were allowed to vary in order to vary $R$. On the contrary, since we are allowing $\Omega$ and $\overline{\beta}$ to vary our this study, the material parameters are being held constant and we see the $\|\mathbb{D}\|$ varying continuously with $\|\mathbb{S}\|$. However, we can show that it is possible to reproduce the results that have been obtained in previous works. For instance, when we set $\epsilon = 2$, $a_1 = 0.2$, $a_2 = 1$, and $\overline{\beta}=0.1$ for $R = 10$, we can obtain the same result as that in Figures 3 and 4 of \cite{fusi2023flow}. (See Figure \ref{Fig33})

\begin{figure}[H]
    \centering
    \includegraphics[scale=0.9]{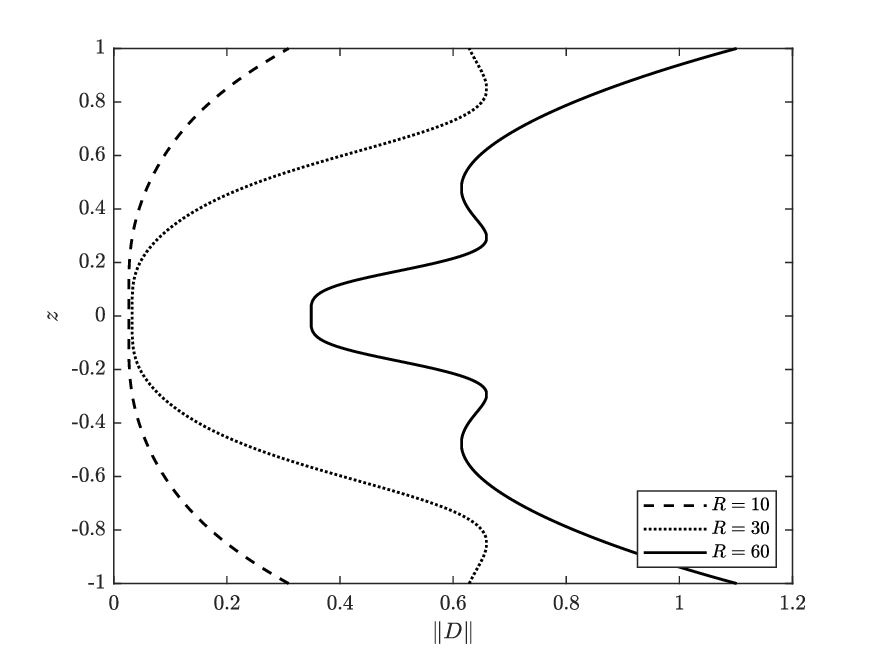}
    \caption{$\|\mathbb{D}\|$ as $R$ varies when $a_1 = 0.05$, $m = -0.8$}
    \label{Fig5.10}
\end{figure}

Similar to the behaviour $\|\mathbb{D}\|$ shows in Figure \ref{Fig5.8} when $a_1 = 0$, here too we see the region with maximum $\|\mathbb{D}\|$ moving closer to the center as $R$ increases. But as $R$ increases further, the region with highest $\|\mathbb{D}\|$ is once again at the plates. This has been shown in Figure \ref{Fig5.10}.

\begin{figure}[H]
    \centering
    \includegraphics[scale=0.9]{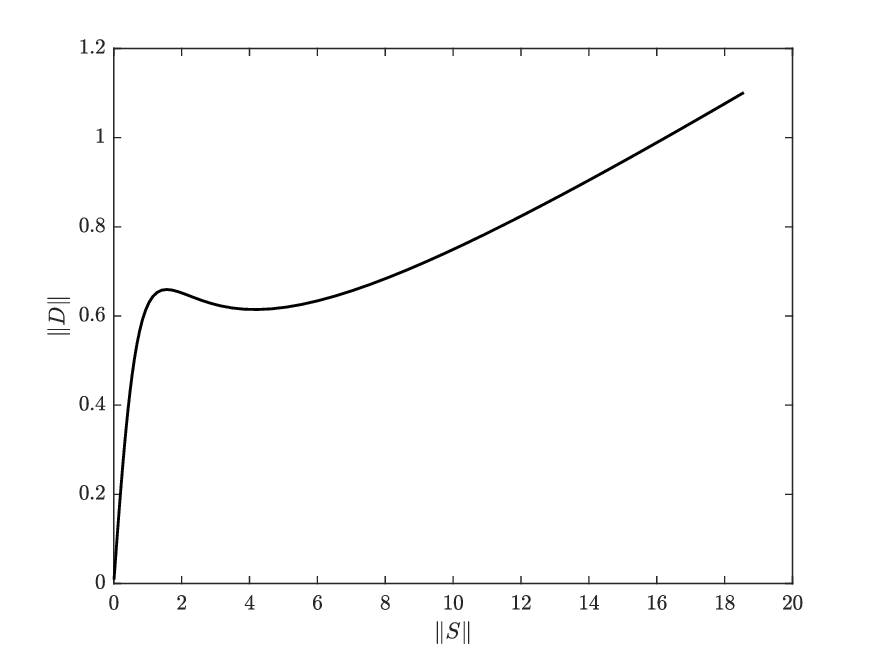}
    \caption{$\|\mathbb{D}\|$ vs $\|\mathbb{S}\|$ at $z = 1$ for $a_1 = 0.05$, $m = -0.8$}
    \label{Fig5.11}
\end{figure}

We can also plot the corresponding values of $\|\mathbb{S}\|$ and $\|\mathbb{D}\|$ at $z=1$ for all the values of $R$ from 1 to 60, and we can see that it resembles the trend expected form our constitutive relation with an S-type non-monotonicity as shown in Figure \ref{Fig5.11}.

\subsection{Stress-limiting case}
\begin{figure}[H]
 \vspace{1pt}
 \centering
% % =================================================
% % First image, top left.
% % =================================================
 \begin{subfigure}{.5\linewidth}
     \centering
    \includegraphics[width=1.05\linewidth]{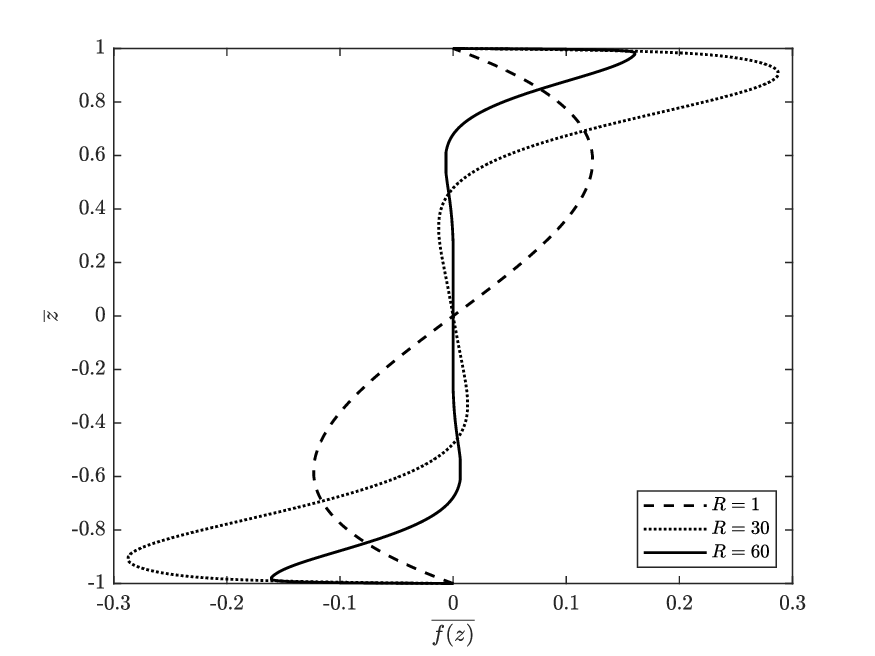}
    %\caption{$f(z)$}
    %\label{Img:1TL}
 \end{subfigure}%\\[-5ex]
% % =================================================
% % Top right
% % =================================================
 \begin{subfigure}{.5\linewidth}
    \centering
    \includegraphics[width=1.05\linewidth]{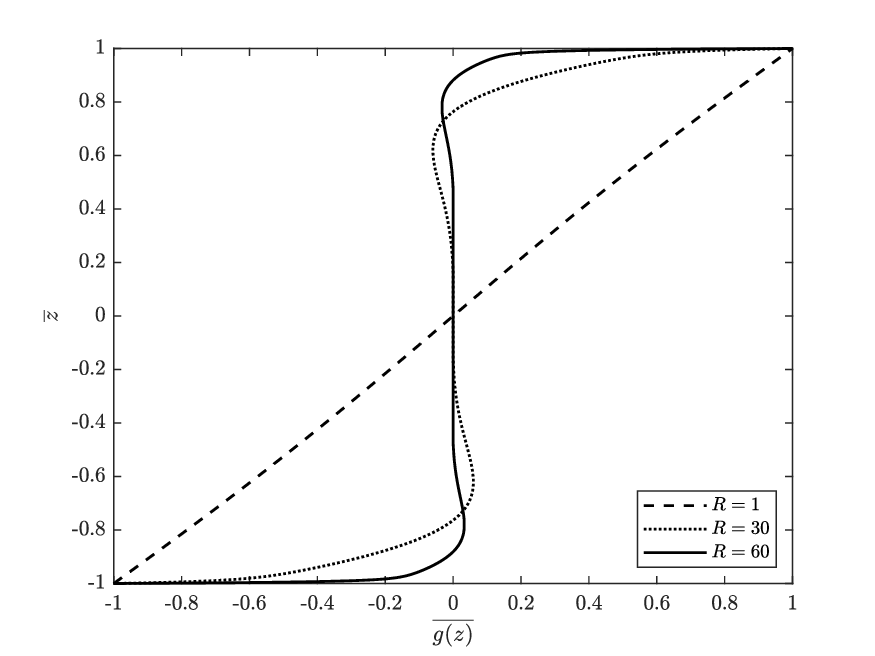}
    %\caption{$g(z)$}
    %\label{Img:1TR}
 \end{subfigure}\\[-5ex]
% % =================================================
% % Bottom left
% % =================================================
 \begin{subfigure}{.5\linewidth}
     \centering
     \includegraphics[width=1.05\linewidth]{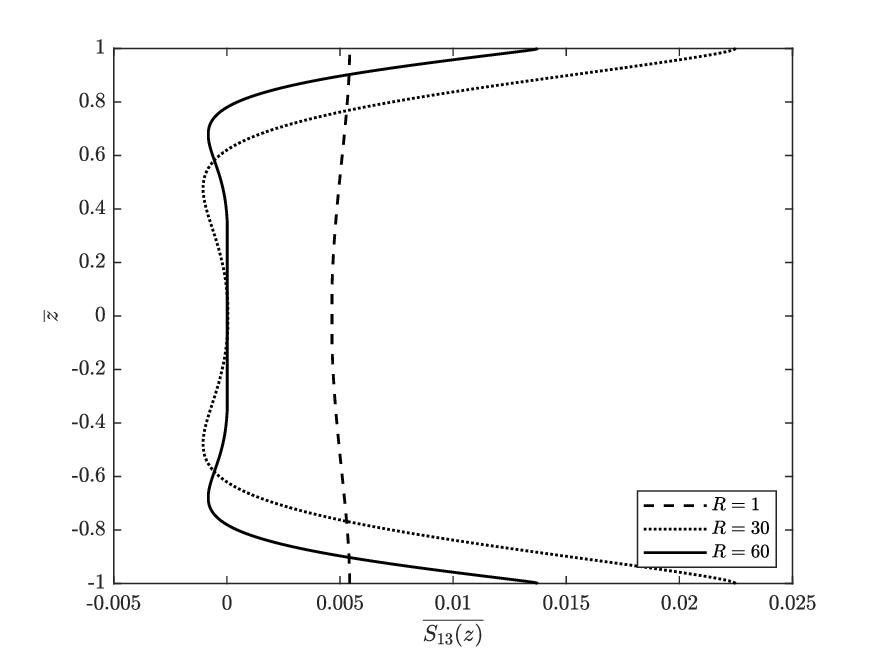}
    %\caption{$S_{13}$}
    %\label{Img:1BL}
 \end{subfigure}%\\[-5ex]
% % =================================================
% % Bottom right
% % =================================================
 \begin{subfigure}{.5\linewidth}
     \centering
     \includegraphics[width=1.05\linewidth]{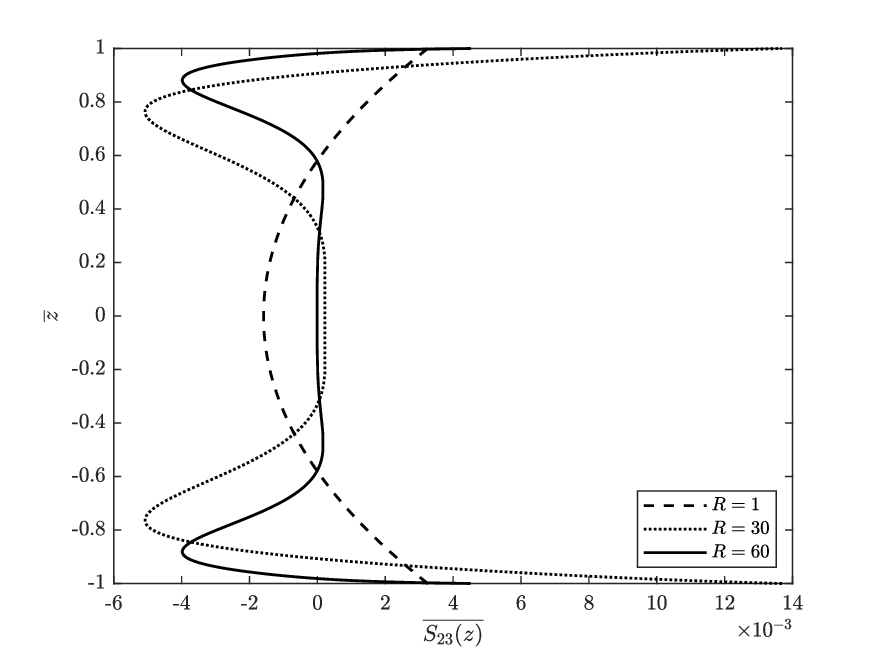}
    %\caption{$S_{23}$}
     %\label{Img:1BR}
 \end{subfigure}%\\[-5ex]
% % =================================================
% % Caption for entire figure
% % =================================================
 \caption{Solution for stress-limiting fluid at various Reynolds numbers for $m = 10$}
 \label{Fig5.12}
 \vspace{1pt}
% % =================================================
\end{figure}

When $a_1 = 1$, $a_2 = 0$ and $m>0$, as $m$ increases, model (\ref{Eqn1}) starts to show stress limiting behaviour. For high values of $m$, when $\|\mathbb{S}\|$ is small, $\|\mathbb{S}\|^{2m}$ is much smaller. This means for lower stresses, linear behaviour dominates and when $\|\mathbb{S}\|>1$, the fluidity drastically increases tending to an Euler fluid as $\|\mathbb{S}\|$ increases. The boundary layers are much sharper (See Figure \ref{Fig5.12}). 

Recently, Garimella et al. \cite{garimella2022new} introduced a two parameter model that mimics the response of a viscoplastic fluid whose viscosity depends exponentially on the shear-rate. Fusi et al. \cite{fusi2023}, in their study of the flow of such a fluid using an orthogonal rheometer, similarly observed the formation of boundary layers, even at small Reynolds numbers.

\subsection{Fluids with large zero-stress fluidity}

\begin{figure}[H]
 \vspace{1pt}
 \centering
% % =================================================
% % First image, top left.
% % =================================================
 \begin{subfigure}{.5\linewidth}
     \centering
    \includegraphics[width=1.05\linewidth]{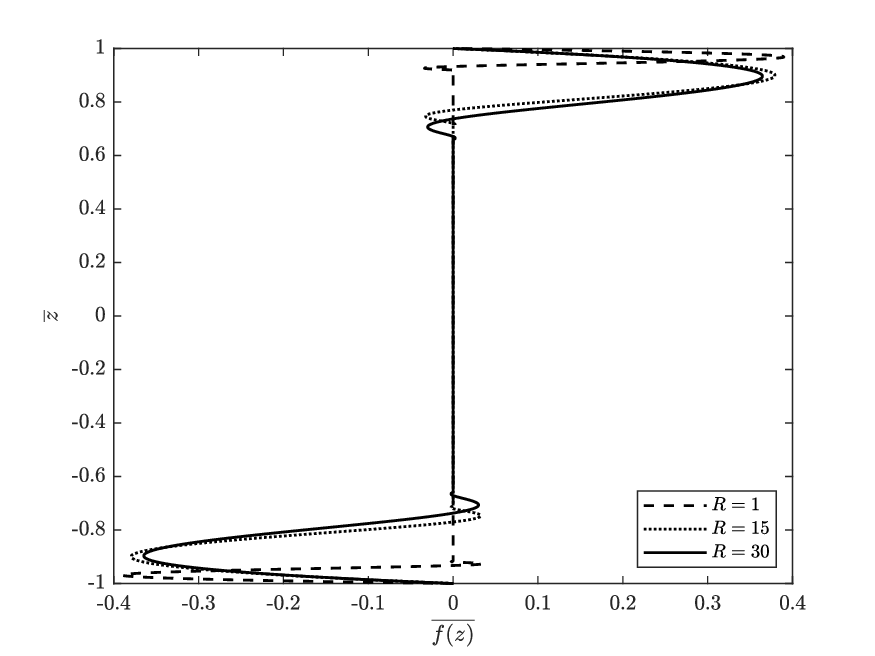}
    %\caption{$f(z)$}
    %\label{Img:1TL}
 \end{subfigure}%\\[-5ex]
% % =================================================
% % Top right
% % =================================================
 \begin{subfigure}{.5\linewidth}
    \centering
    \includegraphics[width=1.05\linewidth]{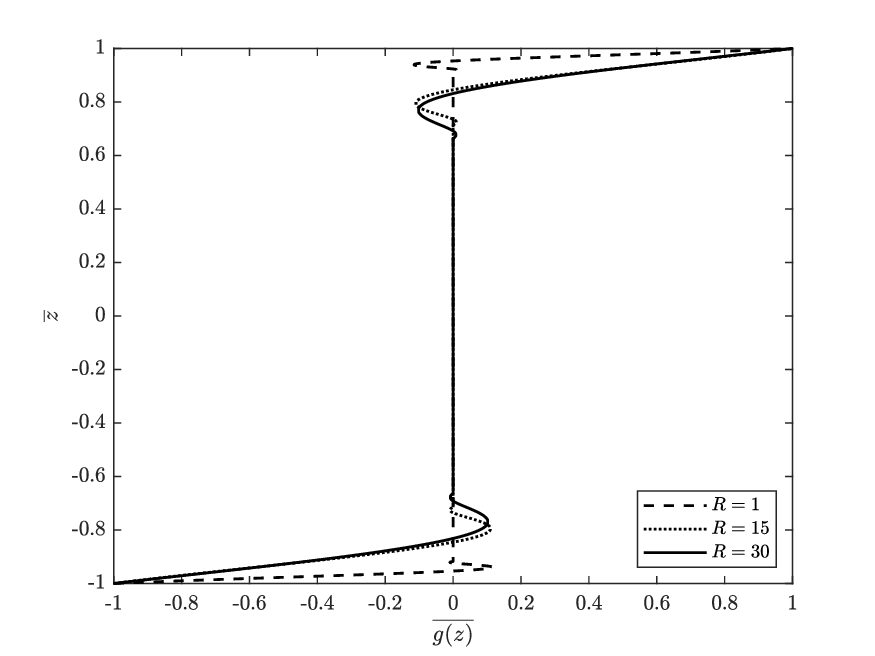}
    %\caption{$g(z)$}
    %\label{Img:1TR}
 \end{subfigure}\\[-5ex]
% % =================================================
% % Bottom left
% % =================================================
 \begin{subfigure}{.5\linewidth}
     \centering
     \includegraphics[width=1.05\linewidth]{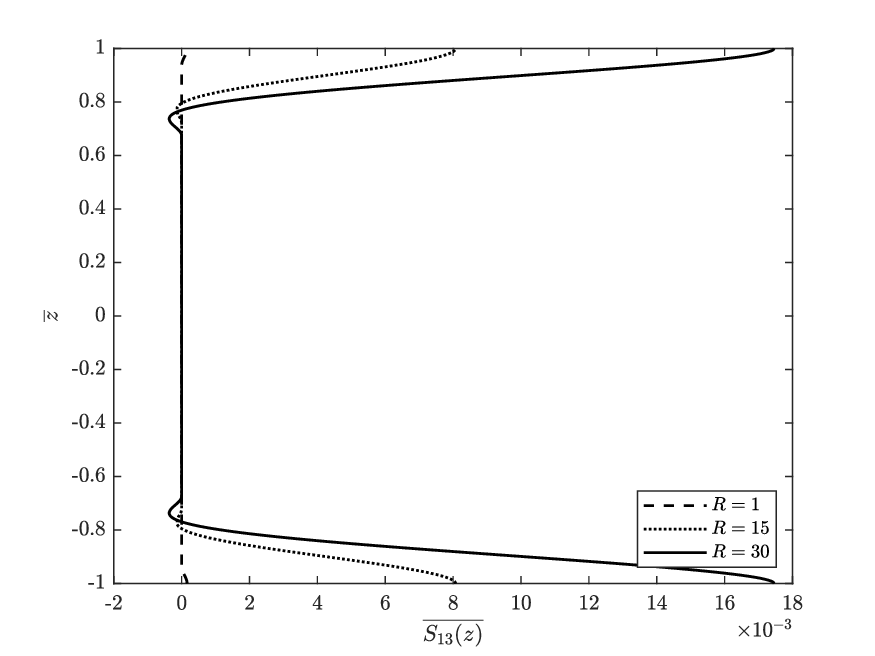}
    %\caption{$S_{13}$}
    %\label{Img:1BL}
 \end{subfigure}%\\[-5ex]
% % =================================================
% % Bottom right
% % =================================================
 \begin{subfigure}{.5\linewidth}
     \centering
     \includegraphics[width=1.05\linewidth]{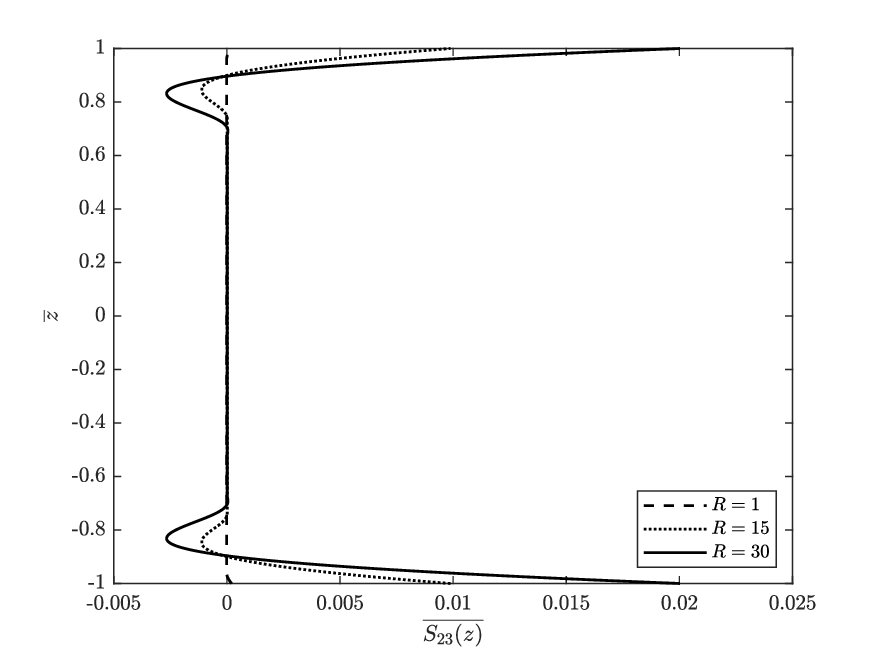}
    %\caption{$S_{23}$}
     %\label{Img:1BR}
 \end{subfigure}%\\[-5ex]
% % =================================================
% % Caption for entire figure
% % =================================================
 \caption{Solution for fluids with very large zero-stress fluidity at various Reynolds numbers for $m = -0.4$}
 \label{Fig5.13}
 \vspace{1pt}
% % =================================================
\end{figure}

If we have $a_1=1$, $a_2 = 0$ and $m<0$ in (\ref{Eqn1}), the model describes fluids whose fluidity becomes very large and behaves like an Euler fluid for very small stress and then as stress increases, it starts to behave like a linearly viscous fluid. The solution we would obtain for $m = -0.4$, $\beta = 1$ and $\epsilon = 0.01$ at various Reynolds numbers is given in Figure \ref{Fig5.13}. We see extremely sharp boundary layers at very small Reynolds numbers and as $R$ increases, we see an increase in the boundary layer thickness.

\subsection{Shear-thickening case}
\begin{figure}[H]
 \vspace{1pt}
 \centering
% % =================================================
% % First image, top left.
% % =================================================
 \begin{subfigure}{.5\linewidth}
     \centering
    \includegraphics[width=1.05\linewidth]{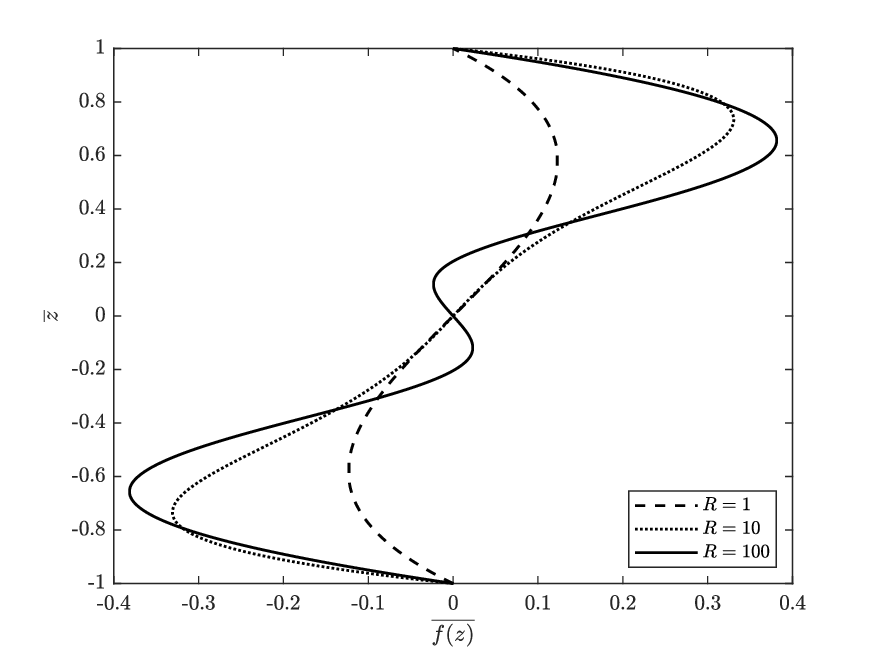}
    %\caption{$f(z)$}
    %\label{Img:1TL}
 \end{subfigure}%\\[-5ex]
% % =================================================
% % Top right
% % =================================================
 \begin{subfigure}{.5\linewidth}
    \centering
    \includegraphics[width=1.05\linewidth]{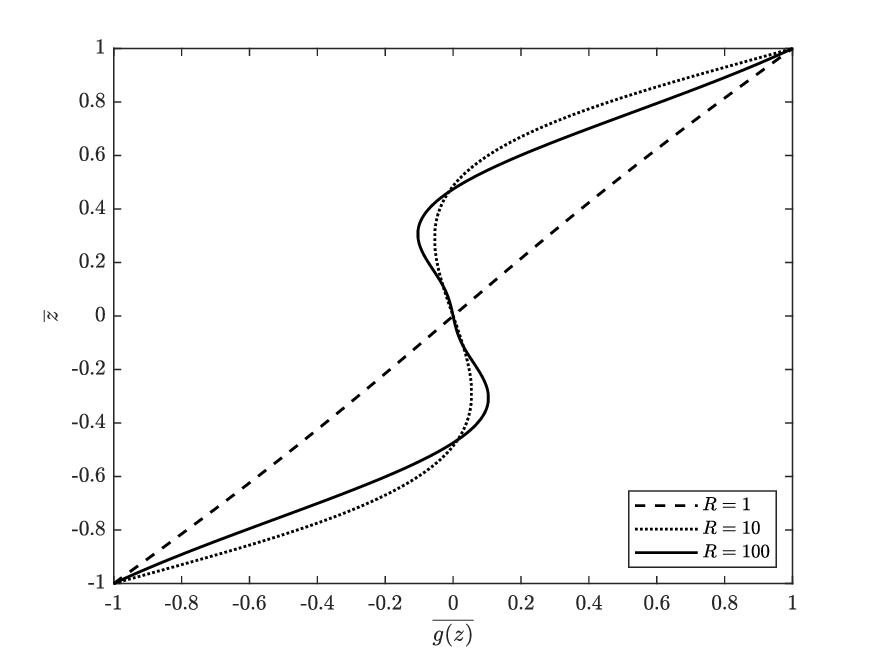}
    %\caption{$g(z)$}
    %\label{Img:1TR}
 \end{subfigure}\\[-5ex]
% % =================================================
% % Bottom left
% % =================================================
 \begin{subfigure}{.5\linewidth}
     \centering
     \includegraphics[width=1.05\linewidth]{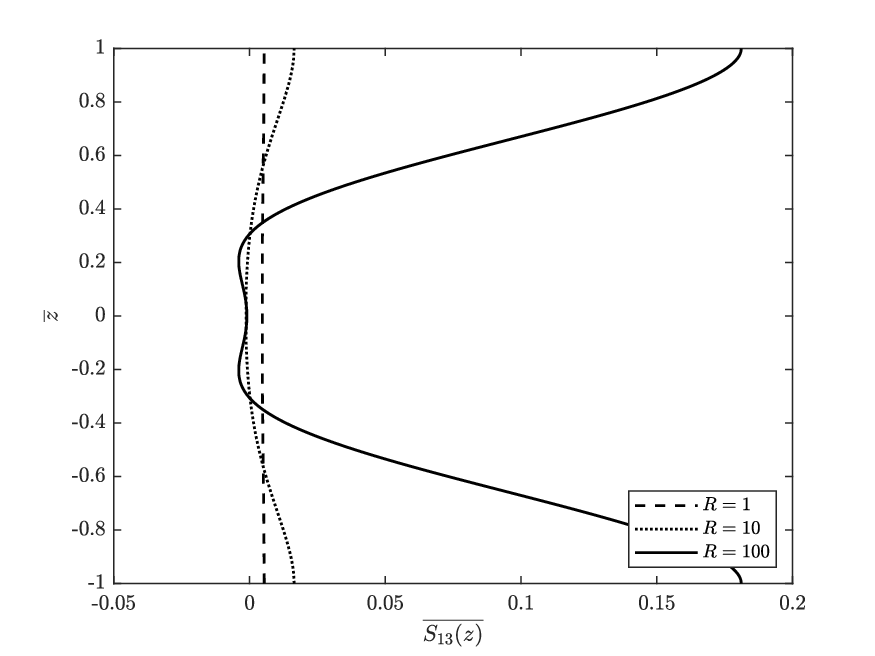}
    %\caption{$S_{13}$}
    %\label{Img:1BL}
 \end{subfigure}%\\[-5ex]
% % =================================================
% % Bottom right
% % =================================================
 \begin{subfigure}{.5\linewidth}
     \centering
     \includegraphics[width=1.05\linewidth]{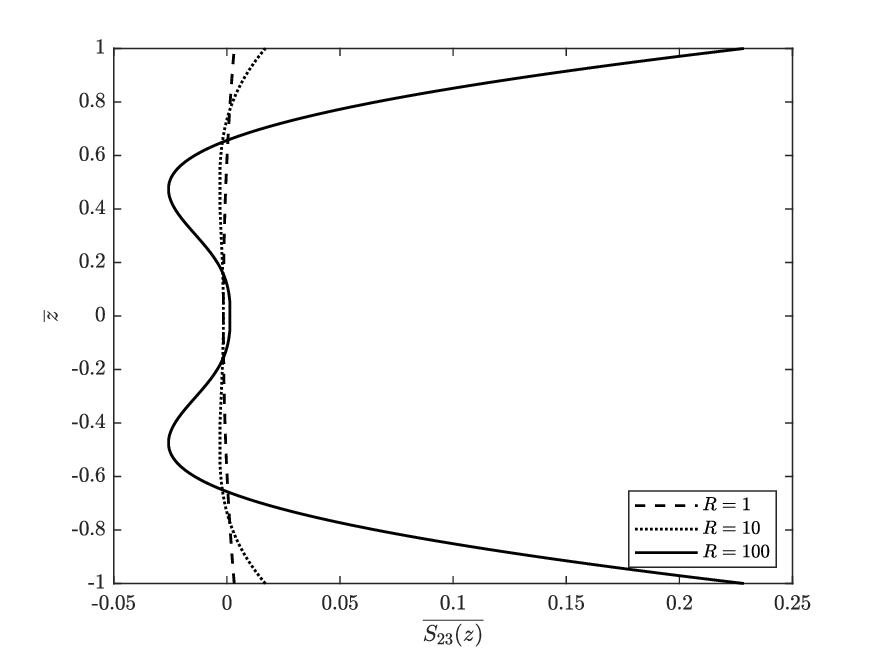}
    %\caption{$S_{23}$}
     %\label{Img:1BR}
 \end{subfigure}%\\[-5ex]
% % =================================================
% % Caption for entire figure
% % =================================================
 \caption{Solution for shear-thickening fluid at various Reynolds numbers for $b_1 = 0$, $b_2 = 1$ and $n = 2$}
 \label{Fig5.14}
 \vspace{1pt}
% % =================================================
\end{figure}

Now if we look at the classical counter-part (\ref{Eqn2}) of the generalized stress-power law model, for $b_2 = 1$ and $n>0$, the model shows shear thickening behaviour. As the shear-rate increases, the viscosity of the fluid also increases. The solution for such a fluid with $b_1 = 0$, $b_2 = 1$ and $n = 2$ is given in Figure \ref{Fig5.14}. We see that the slopes of $\overline{f}$ and $\overline{g}$ with respect to $\overline{z}$ become steeper with increasing $R$ initially and the start to decrease. We do not see the presence of boundary layers even until $R = 100$.

\subsection{Shear-thinning case}
\begin{figure}[H]
 \vspace{1pt}
 \centering
% % =================================================
% % First image, top left.
% % =================================================
 \begin{subfigure}{.5\linewidth}
     \centering
    \includegraphics[width=1.05\linewidth]{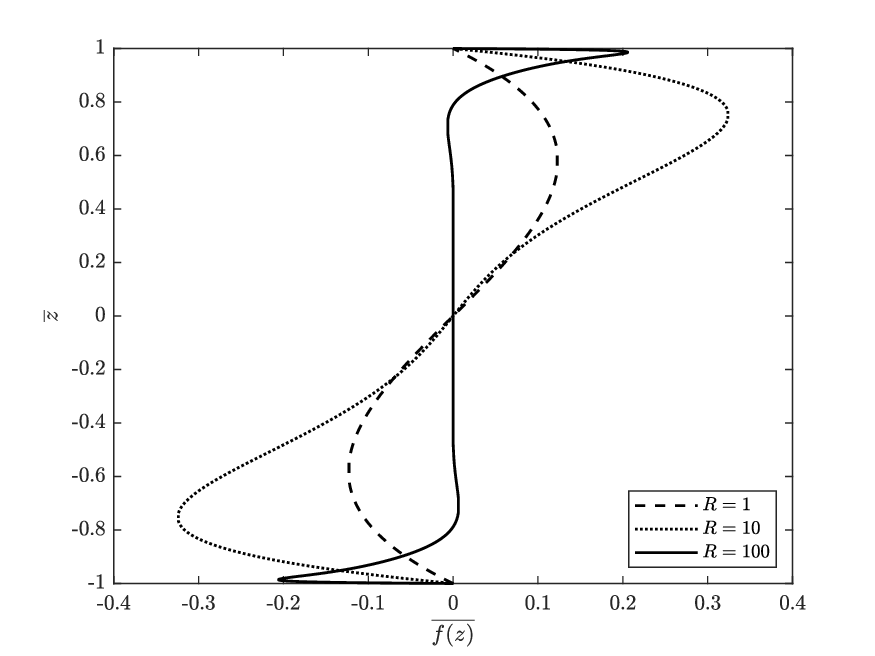}
    %\caption{$f(z)$}
    %\label{Img:1TL}
 \end{subfigure}%\\[-5ex]
% % =================================================
% % Top right
% % =================================================
 \begin{subfigure}{.5\linewidth}
    \centering
    \includegraphics[width=1.05\linewidth]{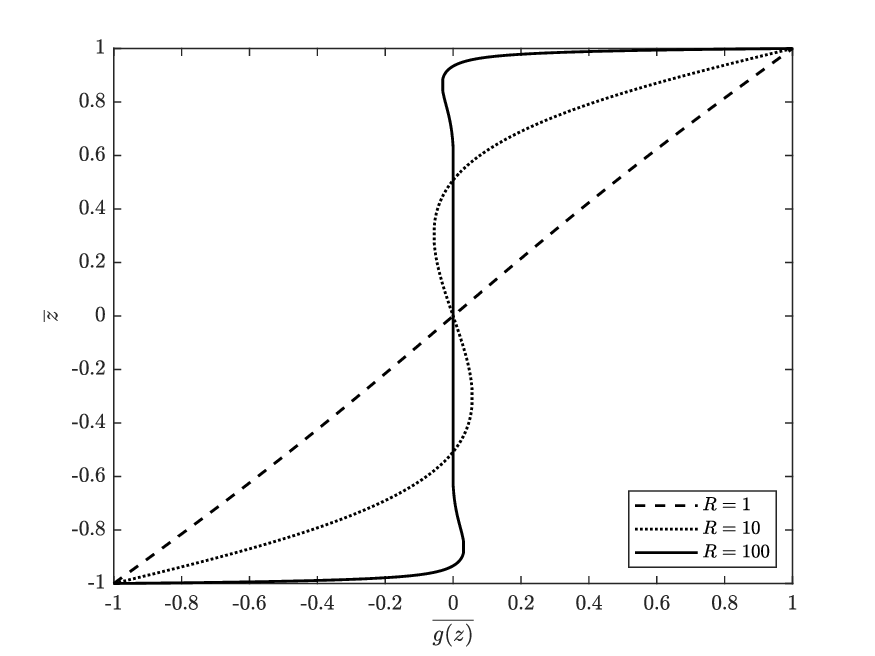}
    %\caption{$g(z)$}
    %\label{Img:1TR}
 \end{subfigure}\\[-5ex]
% % =================================================
% % Bottom left
% % =================================================
 \begin{subfigure}{.5\linewidth}
     \centering
     \includegraphics[width=1.05\linewidth]{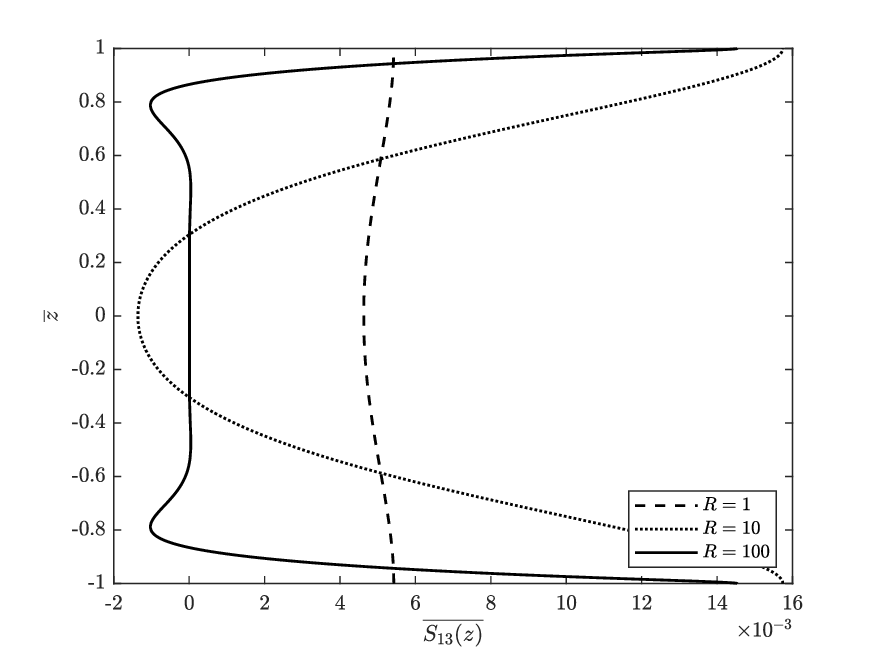}
    %\caption{$S_{13}$}
    %\label{Img:1BL}
 \end{subfigure}%\\[-5ex]
% % =================================================
% % Bottom right
% % =================================================
 \begin{subfigure}{.5\linewidth}
     \centering
     \includegraphics[width=1.05\linewidth]{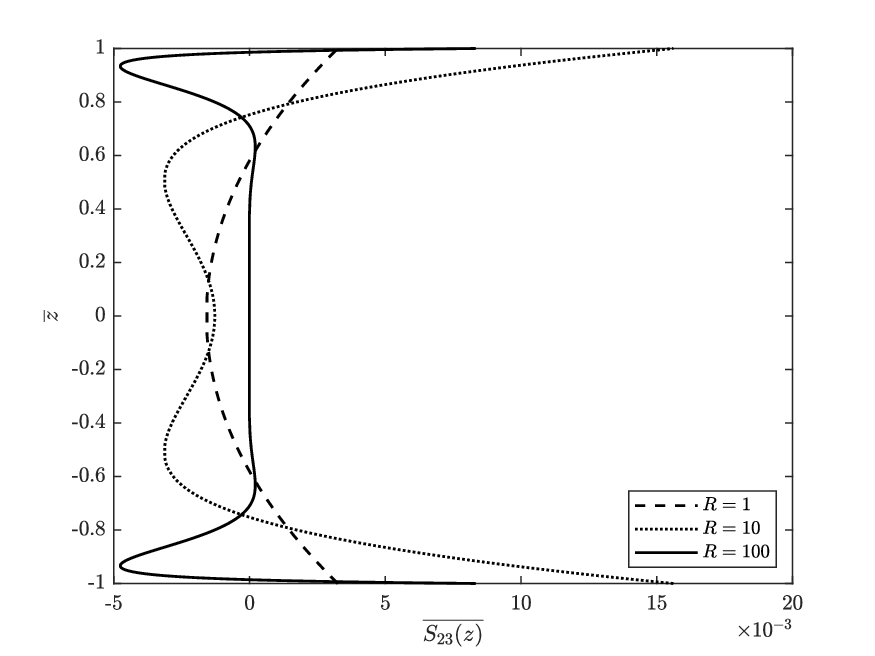}
    %\caption{$S_{23}$}
     %\label{Img:1BR}
 \end{subfigure}%\\[-5ex]
% % =================================================
% % Caption for entire figure
% % =================================================
 \caption{Solution for shear-thinning fluid at various Reynolds numbers for $b_1 = 0$, $b_2 = 1$ and $n = -0.4$}
 \label{Fig5.15}
 \vspace{1pt}
% % =================================================
\end{figure}

In (\ref{Eqn2}), for $b_2 = 0$ and $n<0$, the fluid exhibits shear-thinning behaviour. We can see from the solution given in Figure \ref{Fig5.15} that the boundary layer reduces as $R$ increases and the stresses are also much lower compared to the shear-thickening case for the same Reynolds numbers, which is expected. 

\subsection{Shear-limiting case}
\begin{figure}[H]
 \vspace{1pt}
 \centering
% % =================================================
% % First image, top left.
% % =================================================
 \begin{subfigure}{.5\linewidth}
     \centering
    \includegraphics[width=1.05\linewidth]{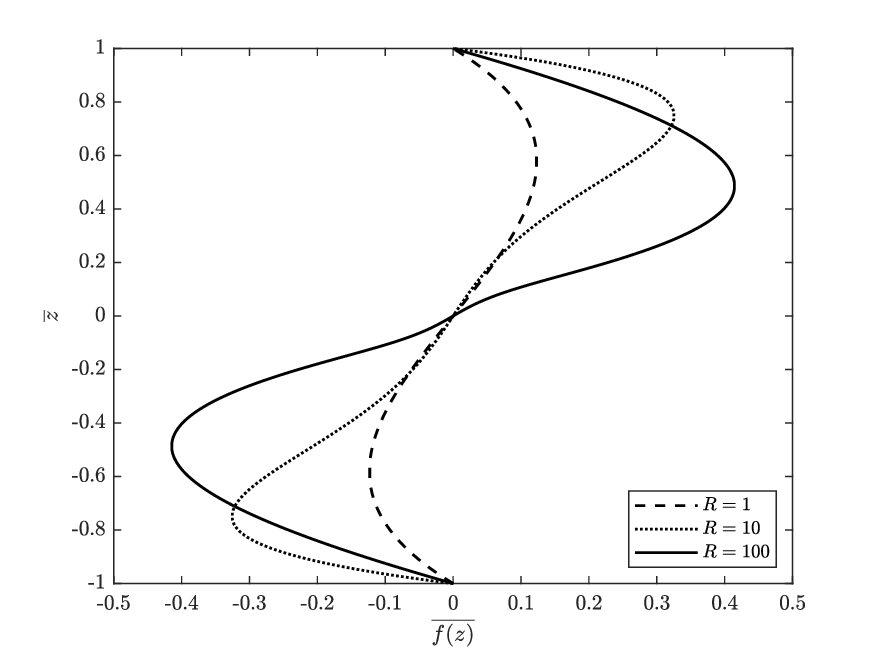}
    %\caption{$f(z)$}
    %\label{Img:1TL}
 \end{subfigure}%\\[-5ex]
% % =================================================
% % Top right
% % =================================================
 \begin{subfigure}{.5\linewidth}
    \centering
    \includegraphics[width=1.05\linewidth]{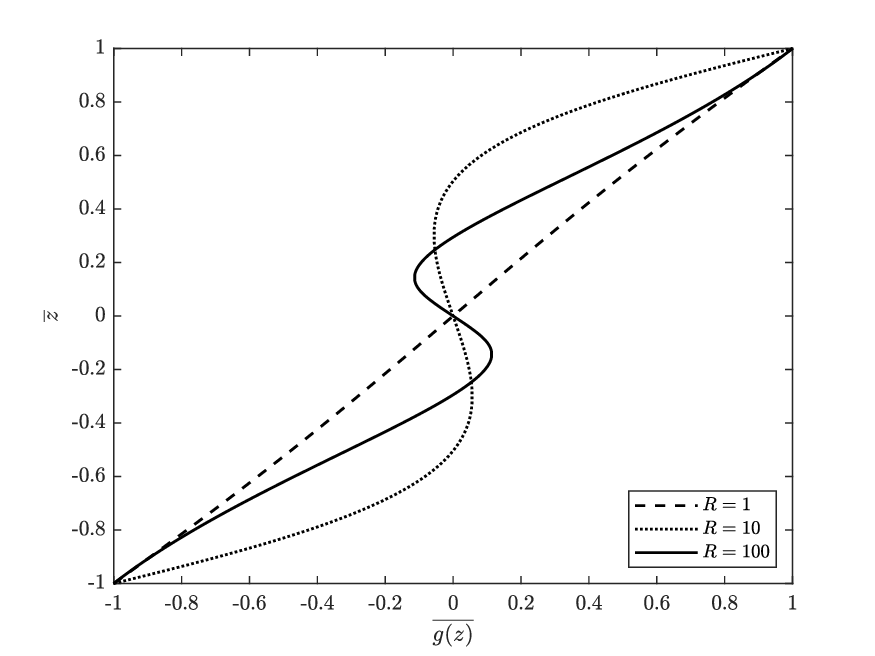}
    %\caption{$g(z)$}
    %\label{Img:1TR}
 \end{subfigure}\\[-5ex]
% % =================================================
% % Bottom left
% % =================================================
 \begin{subfigure}{.5\linewidth}
     \centering
     \includegraphics[width=1.05\linewidth]{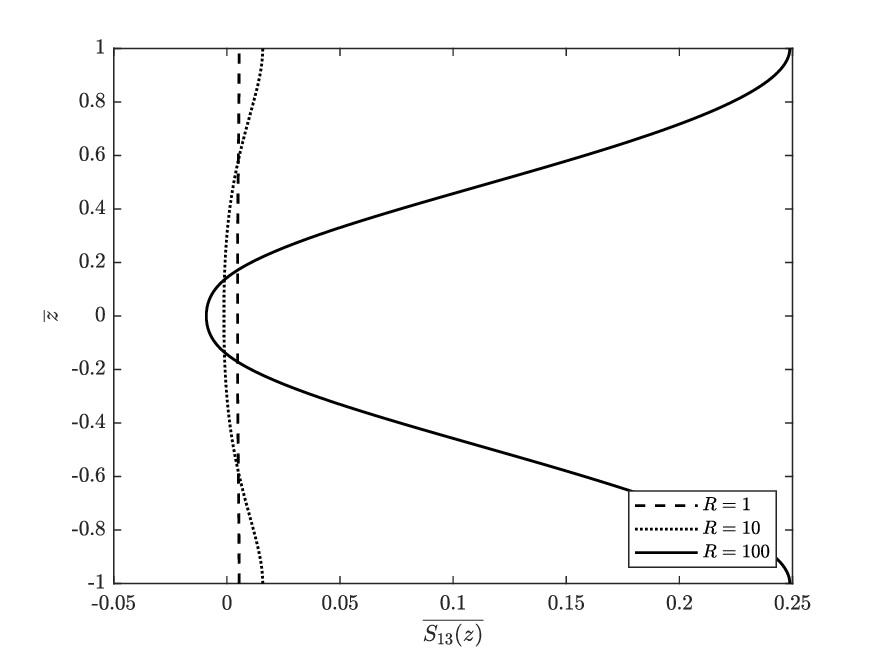}
    %\caption{$S_{13}$}
    %\label{Img:1BL}
 \end{subfigure}%\\[-5ex]
% % =================================================
% % Bottom right
% % =================================================
 \begin{subfigure}{.5\linewidth}
     \centering
     \includegraphics[width=1.05\linewidth]{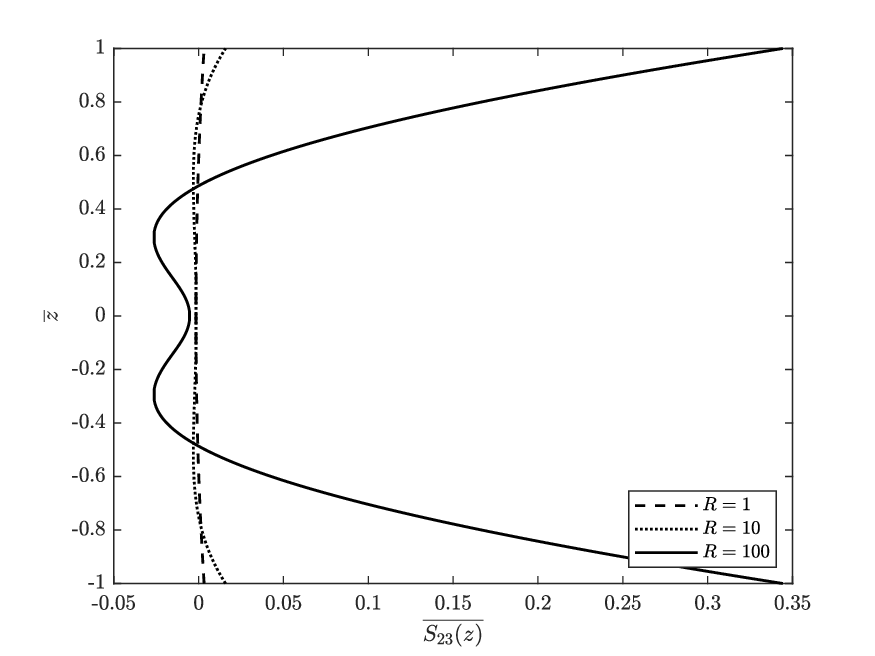}
    %\caption{$S_{23}$}
     %\label{Img:1BR}
 \end{subfigure}%\\[-5ex]
% % =================================================
% % Caption for entire figure
% % =================================================
 \caption{Solution for fluid that shows shear limiting behaviour at various Reynolds numbers for $b_1 = 1$, $b_2 = 0$ and $n = 10$}
 \label{Fig5.16}
 \vspace{1pt}
% % =================================================
\end{figure}

When in (\ref{Eqn2}) we have, $b_1 = 1$, $b_2 = 0$, and when $n$ is a large positive number, for small values of $\|\mathbb{D}\|$, the model initially exhibits linearly viscous behavior and with increasing $\|\mathbb{D}\|$, the viscosity drastically increases resulting in shear-limiting behaviour. The solution for such a model with $n = 10$ for various Reynolds numbers is given in Figure \ref{Fig5.16}.

\subsection{Fluids with large zero-shear viscosity}
\begin{figure}[H]
 \vspace{1pt}
 \centering
% % =================================================
% % First image, top left.
% % =================================================
 \begin{subfigure}{.5\linewidth}
     \centering
    \includegraphics[width=1.05\linewidth]{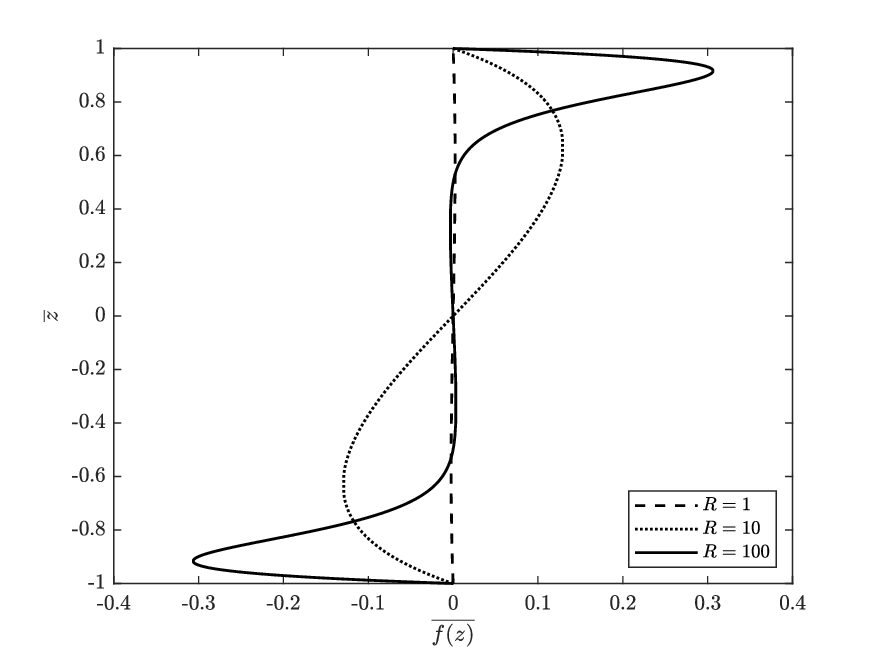}
    %\caption{$f(z)$}
    %\label{Img:1TL}
 \end{subfigure}%\\[-5ex]
% % =================================================
% % Top right
% % =================================================
 \begin{subfigure}{.5\linewidth}
    \centering
    \includegraphics[width=1.05\linewidth]{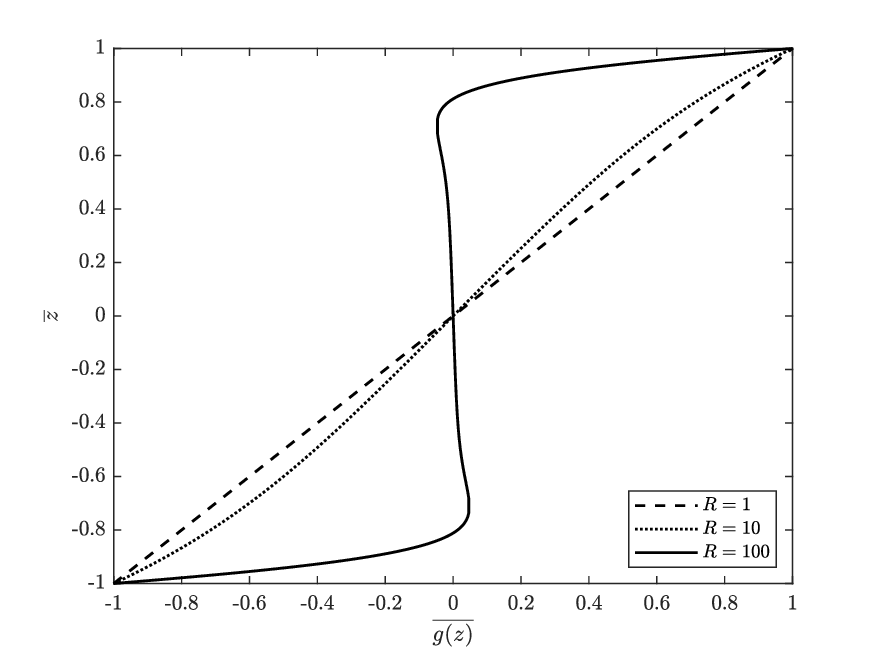}
    %\caption{$g(z)$}
    %\label{Img:1TR}
 \end{subfigure}\\[-5ex]
% % =================================================
% % Bottom left
% % =================================================
 \begin{subfigure}{.5\linewidth}
     \centering
     \includegraphics[width=1.05\linewidth]{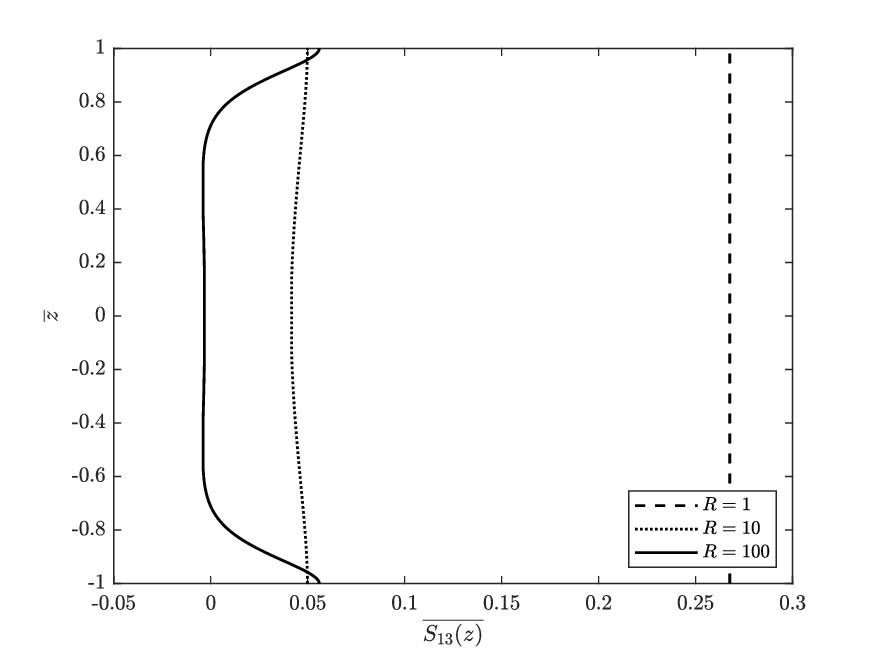}
    %\caption{$S_{13}$}
    %\label{Img:1BL}
 \end{subfigure}%\\[-5ex]
% % =================================================
% % Bottom right
% % =================================================
 \begin{subfigure}{.5\linewidth}
     \centering
     \includegraphics[width=1.05\linewidth]{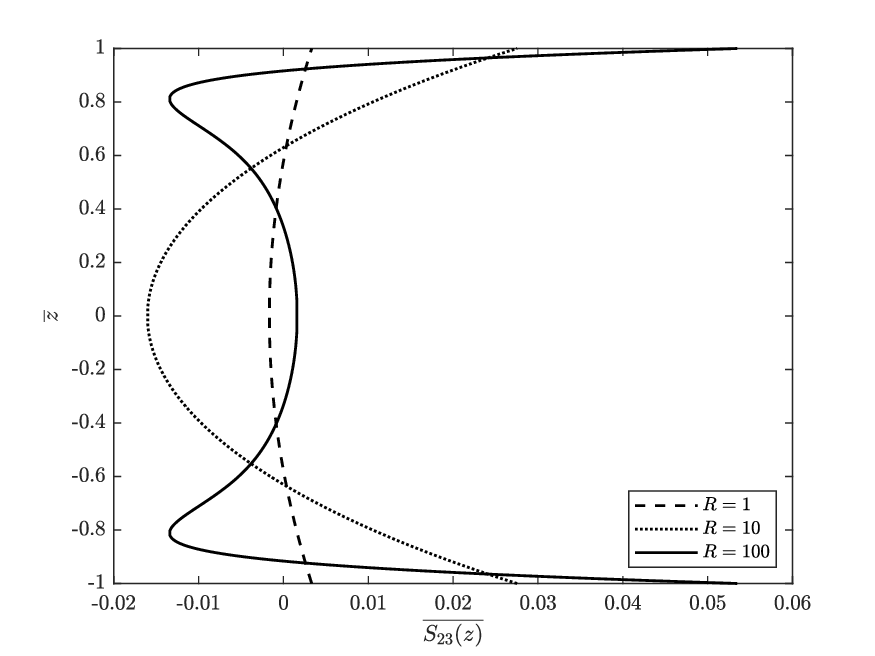}
    %\caption{$S_{23}$}
     %\label{Img:1BR}
 \end{subfigure}%\\[-5ex]
% % =================================================
% % Caption for entire figure
% % =================================================
 \caption{Solution for fluid that shows large zero-shear viscosity at various Reynolds numbers for $b_1 = 1$, $b_2 = 0$ and $n = -0.4$}
 \label{Fig5.17}
 \vspace{1pt}
% % =================================================
\end{figure}

When in (\ref{Eqn2}) we have, $b_1 = 1$, $b_2 = 0$ and $-0.5<n<0$, the fluid shows infinite zero-shear viscosity like leading to a ``yield-stress" as $\|\mathbb{D}\| \rightarrow 0$, and as $\|\mathbb{D}\|$ increases the fluid starts to exhibit linearly viscous behaviour. Solution for stresses and the projections of the locus of centers of rotation for such a fluid is as shown in Figure \ref{Fig5.17}.

\noindent
\textbf{Acknowledgement}

\noindent
This research was funded by Texas A\&M Engineering Experiment Section (TEES).

\bibliographystyle{elsarticle-num}
\bibliography{references}
\end{document}